\documentclass{aa}
\usepackage{graphicx}
\usepackage{lscape}
\usepackage{longtable}

\usepackage{txfonts}
\def\Msun{M$_{\sun}$}
\def\Mjup{M$_{\rm Jup}$}
\def\ms{m\,s$^{\rm -1}$}
\def\kms{km\,s$^{\rm -1}$}
\def\vsini{$v\sin{i}$}
\def\ie{{\it i.e.}, }
\def\eg{{\it e.g.}, }
\def\bv{$B-V$}

\begin{document}
   \title{Extrasolar planets and brown dwarfs around A--F type stars
     \thanks{Based on observations collected at the European Southern Observatory, Chile, ESO 075.C-0689, 076.C-0279, 077.C-0295, 078.C-0209, 080.C-0664, 080.C-0712.}}

 \subtitle{VI. High precision RV survey of early type dwarfs with HARPS}

   \author{
     A.-M. Lagrange \inst{1}
     \and
     M. Desort \inst{1}
     \and
     F. Galland \inst{1}
     \and
     S. Udry \inst{2}
     \and
     M. Mayor \inst{2}
    }

   \offprints{
     A.-M. Lagrange,\\
     \email{anne-marie.lagrange@obs.ujf-grenoble.fr}
   }

   \institute{
     Laboratoire d'Astrophysique de l'Observatoire de Grenoble,
     Universit\'e Joseph Fourier, BP 53, 38041 Grenoble, France
     \and
     Observatoire de Gen\`eve, 51 Ch. des Maillettes, 1290 Sauverny,
     Switzerland
    }

   \date{Received date / Accepted date}

   
   \abstract
   {}
   {Systematic surveys to search for exoplanets have been mostly dedicated to solar-type stars sofar. We developed in 2004 a method to extend such searches to earlier A--F type dwarfs and started spectroscopic surveys to search for planets and quantify the detection limit achievable when taking into account the stars properties (Spectral Type, \vsini) and their actual levels of intrinsic variations. We give here the first results of our southern survey with {\small HARPS}.
   }
   {We observed 185 A--F (\bv~in the range [$-$0.1; 0.6]) stars with {\small HARPS} and analysed them with our dedicated software. We use several criteria to probe different origins for the radial-velocity variations -- stellar activity (spots, pulsations) or companions: bisector shape, radial-velocity variations amplitudes and timescales.
   }
   {1) 64\,$\%$ of the 170 stars with enough data points are found to be variable. 20 are found to be binaries or candidate binaries (with stars or brown dwarfs). More than 80\,$\%$ or the latest type stars (once binaries are removed) are intrinsically variable at a 2\,\ms precision level. Stars with earlier spectral type (\bv~$\leq$ 0.2) are either variable or associated to levels of uncertainties comparable to the RV rms observed on variable stars of same \bv. 2) We have detected one long-period planetary system (presented in another paper) around an F6IV--V star. 3) We have quantified the jitter due to stellar activity and we show that taking into account this jitter in addition to the stellar parameters (spectral type, \vsini), it is still possible to detect planets with {\small HARPS} with periods of 3 days (resp. 10 days and 100 days) on 91\,$\%$ (resp. 83\,$\%$, 61\,$\%$) of them. We show that even the earliest spectral type stars are accessible to this type of search!
 , provided they have a low projected rotational velocity and low levels of activity. 4) Taking into account the present data, we compute the actually achieved detection limits for 107 targets and discuss the limits as a function of \bv. Given the data at hand, our survey is sensitive to short-period (few days) planets and to longer ones (100 days) at a lower extent (latest type stars). We derive first constrains on the presence of planets around A--F stars for these ranges of periods.  
   }
   {} 

   \maketitle
%

\section{Introduction}

Since the discovery of the first exoplanet around a solar-like star in 1995, more than 250 planets have been found by radial-velocity (RV) surveys (Jean Schneider, http://exoplanet.eu). These surveys have generally focused on late-type stars (later than F8). However, knowing about the presence of planets or brown dwarfs (hereafter BDs) around more massive objects is mandatory if one wishes to investigate the impact of the mass of the central stars on the planetary formation and evolution processes.

There are theoretical indications that the mass of the planets increases with the mass of the parent star, at least for low mass stars (\cite{ida05}) and that the frequency of giant planets increases linearly with the parent star mass for stars between 0.4 and 3\,\Msun~(\cite{kennedy08}), with \eg 6\,$\%$ frequency of giant planets around 1\,\Msun~and 10\,$\%$ frequency around 1.5\,\Msun. More numerous and massive planets are consistent with what we could expect from a disk surface density increasing with stellar mass. On the other hand the shorter lifetimes of the systems as well as the lack of solid material close to the star could reduce the number of planets. Clearly, several parameters probably impact the occurence and properties of planets around massive stars, and they have not yet been fully explored.

The data to test the models are still quite limited as the largest and earliest, now long-lasting surveys had focused on solar type, main-sequence (MS) stars. In recent years, some efforts have been made nevertheless to search for planets around stars with various masses: less massive, M-type stars on the one hand, and more massive stars on the other hand. The search for planets around M stars seems sofar to confirm the previously mentionned expectations from theoretical works (see \eg \cite{bonfils05}; \cite{butler06}). As far as massive stars are concerned, the available observations are still very limited. Massive MS stars have been removed from early surveys as it was generally thought that their spectra (few lines, usually broadened by stellar rotation) would not allow planet detection and indeed, the classical RV measurements technics (based on the cross correlation of the actual spectra with a binary spectral mask corresponding to a star with an appropriate spectral type and \vsini~= 0\,\kms) fail to measure the RV of these stars. This lead some groups to study instead ``retired'' early-type, either low-mass ($\leq$ 1.6\,\Msun) giants, intermediate-mass (1.6--2\,\Msun) subgiants or clump giants (1.7--3.9\,\Msun) (see \eg {\bf \cite{hatzes05}}; {\bf \cite{niedzielski07}}; {\bf \cite{johnson06}}; {\bf \cite{johnson07}}; \cite{lovis07}; \cite{sato08}). These stars indeed have cooled out and also rotate more slowly due to coupling of stellar winds and magnetic fields; they therefore exhibit more numerous, narrower lines, which is adequate for classical RV measurements technics, and their level of activity (jitter) is relatively low (10--20 m\,s$^{\rm -1}$; \cite{hekker06}) for giants and 10\,\ms~for subgiants (\cite{johnson07} and ref. therein; \cite{sato08}). The data available today are still limited compared to those available on solar-type stars and less than 20 planets have been found sofar in total around these evolved stars. So far, the planets found around K subgiants stars with M $\geq$ 1.5\,\Msun~are located at distances larger than {\bf 0.8}\,AU (\cite{johnson07}); this led these authors to conclude that close-in planets are rare, in agreement with some theoretical predictions on disks depletion timescales (\cite{burkert07}). However, the impact of the post MS evolution of the stars on closer-in planets has not been explored yet for these stars. Concerning giant stars, all planets found so far have relatively long periods, the closest ones beeing reported at less than 0.7\,AU from 2 giants, in addition to a previously reported planet at 0.7\,AU by \cite{sato03}. Numerical simulations by the same authors suggest that planets with orbits inside 0.5--1\,AU around 2--3\,\Msun~stars could be engulfed by the central stars at the tip of RGB due to tidal torque from the central stars. According to them, if one assumes then that most of the clump giants are post RGB stars, there is then a risk that closer planets, if present before had disappeared since the star evolved. In summary, eventhough there are some hints that hot Jupiters are not present around retired stars, it is recognized that data are still needed to definitely confirm this point. Also, as stellar evolutionary processes may have impacted the presence of planets close to the stars, it is acknowledged that data are needed on A--F MS stars (see eg \cite{burkert07}; see also \cite{li08}). We note finally that short-period planets have indeed been found around F5--F6 MS stars through transit observations (see http://exoplanet.eu).

We developed a few years ago a dedicated software to extract the RV data around early type MS stars. The method consists in correlating, in the Fourier space, each spectrum and a reference spectrum built by summing-up all the available spectra for this star. We had shown earlier that with this approach, and taking into account the stars \bv~and projected rotational velocities, it is possible to find planets around A--F type stars (\cite{galland05a}). However, the price to pay is that more measurements are needed to find planets around A and early F dwarfs than for late F and G--K dwarfs, because of the relatively higher uncertainties in the RV measurements due to higher \vsini~and higher effective temperature, and also because of the possible presence of pulsations or spots in the case of late F stars, the impact of which has not been quantified so far. (Note that spots or pulsations become also a limiting factor in the case of later-type stars, as well as pulsations, if one looks for low mass planets.) We started then systematic searches for low-mass companions to A--F type stars, with {\small HARPS} 
in  the southern hemisphere and with {\small ELODIE} and then {\small SOPHIE} at OHP in the northern hemisphere. We found so far a 9.1\,\Mjup~(minimum mass) planet orbiting ($a$ = 1.1\,AU) an F6V star, with \vsini~= 12\,\kms~(\cite{galland05b}). Very promisingly, we also detected a 21\,\Mjup~brown dwarf orbiting ($a$ = 0.2\,AU) a pulsating A9V star with \vsini~= 50\,\kms~(\cite{galland06}); noticeably, in that case, we could disentangle stellar pulsations from the presence of a low-mass companion.

In parallel, we developed detailed simulations of stellar activity (spots) in order to estimate more quantitatively than what was available so far (\cite{saar97}; \cite{hatzes02}) the impact of such stellar activity on RV data and other observables (bisectors, bisectors velocity-span, photometry). We have showed that if the star \vsini~is smaller than the spectrograph spectral resolution, depending on their location with respect to line of sight, depending also on their size, spots with realistic sizes can produce RV variations and bisector velocity-span variations quite similar to those of low mass planets. Hence, low amplitude (level of typically 20\,\ms~or less) planetary-like RV and bisector velocity span variations cannot alone definitely prove the presence of planets around low \vsini~G--K stars (\cite{desort07}) and additional criteria are mandatory to rule out spots: photometry, activity evaluation down to levels relevent to explain the amplitude of RV variations, precise knowledge of the star rotaional period, etc. The situation in that respect is much more favorable in the case of earlier-type stars as they rotate statistically faster and hence the bisector criteria can apply.

The present paper is devoted to our southern hemisphere survey. The sample, observations, measurements and diagnostics are provided in Sect.~\ref{observ}. The results concerning the stellar variability as well as the quantitative impact on planet detectability around the early-type stars are presented in Sect.~\ref{results}. Finally we give and discuss in Sect.~\ref{detectionlim} the detection limits obtained in the present survey.

\section{Sample, observations and measurements}
\label{observ}

 \subsection{Sample}

 Our {\small HARPS} sample is limited to B8 to F7 dwarfs. The limit in spectral type (ST) at F7 is set because the surveys using masking technique generally start with stars with ST later than F8. The limit at B8 is set by the precision that can be obtained with our method on stars given their ST and their \vsini~(\cite{galland05a}): the detection limit of stars with ST earlier than B8 does not fall into the planet domain. Our survey is also volume-limited with an upper limit at 67\,pc for the B8--A9 dwarfs and at 33\,pc for the F0--F7 dwarfs. The distance was taken from Hipparcos catalogue, and stars with distance uncertainties larger than 20\,$\%$ were removed. The difference in distance for both spectral types comes from the fact that we wanted to have roughly the same number of A and F stars. The dwarf nature was selected by selecting stars with absolute magnitudes within 2.5 magnitude from the Main Sequence.
 
 Spectroscopic binaries as well as close visual binaries with separations smaller than 5\arcsec~known at the begining of the survey from Coravel or Hipparcos data were removed. Confirmed $\delta$\,Scuti (from \cite{rodriguez00}) and $\gamma$\,Doradus type stars (from \cite{mathias04} and http://astro.univie.ac.at/dsn/gerald/gdorlist.html) were also removed as they are known to produce RV variations over hours to a few days periods due to pulsations. Finally we removed as well Ap and Am stars, which present spectral anomalies and are often associated to binary systems. This removes a number of late A - early F type stars, crossing the $\delta$\,Scuti and $\gamma$\,Doradus instability strip. We ended up with 207 stars with ST between B8 and F7, and \bv~ranging respectively between $-$0.1 and 0.58, corresponding to mass ranging between 1.3 and 3.5\,\Msun. Note that we have a relatively smaller number of stars in the [0.2; 0.4] \bv~range (\ie roughly, between 1.8 and 1.4\,\Msun) as we removed the known $\delta$\,Scuti and $\gamma$\,Doradus stars).

 \subsection{Observations}
\begin{figure}[ht]
  \centering
  \includegraphics[angle=0,width=0.95\hsize]{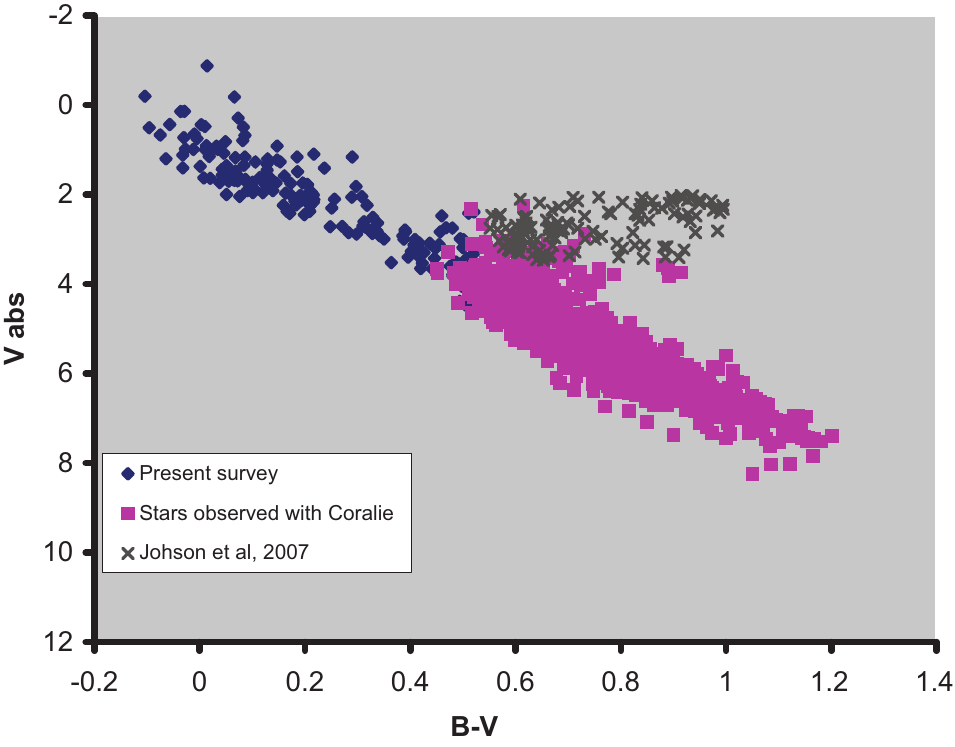}
  \caption{Observed stars in a HR diagram. We also plotted the dwarfs and/or (sub-) giants surveyed either with the Coralie spectrograph or by {\bf Johnson et al, 2006}. Our targets cover a domain that was not surveyed yet. Note the relative lack of objects in the [0.2; 0.4] \bv~region, due to selection effects (see text)}
  \label{targets}
\end{figure}

 185 stars have been observed between August 2005 and January 2008. Figure~\ref{targets} shows their position in the HR diagram. It can be seen that our survey fills a domain of the HR diagram that was not covered yet.

We usually recorded 2 consecutive high resolution ($R \simeq 115000$) spectra each time we pointed to the star (each pointing is hereafter refered to as one epoch). The spectra cover a wavelength range between 3800 and 6900\,$\AA$. As far as possible we moreover tried for a given object to record data at two or three different times during one night, in order to identify possible high-frequency RV variations. We also tried whenever possible to record data on two or three consecutive nights. The time baseline for a given star varies between 5 days and more than 800 days. Only a few (11) stars have been observed during one night only, but for 15 stars we got only 4 good quality (\ie with an absorption\footnote{Magnitude difference between the observation and the one that gives the best signal-to-noise ratio (SN): $abs(i) = 6\log_{10}[ \epsilon_{{\rm rv}}(i) / \min(\epsilon_{{\rm rv}}) ]$, where $\epsilon_{{\rm rv}} (i)$ stands for the uncertainty associated to the measurement of the observation (i) for the considered object, and $min(\epsilon_{{\rm rv}})$ is the smallest value of uncertainty obtained for this object. The measured uncertainties (cf \cite{galland05a}) take into account the photon noise + instrumental uncertainties.} smaller than 2) spectra or less.
We ended up with 170 dwarfs for which we recorded 6 or more good quality spectra. 45 have \bv~$\leq$ 0.1; 72 have \bv~between 0.1 and 0.4 and 53 have \bv~$\geq$ 0.4. We will restrain hereafter our study to those stars.

Typical exposure times ranged between 30\,s and 15\,min depending on the star magnitude and on the atmospheric conditions.

Table\,\ref{sample} provides the 170 targets observed, together with a number of relevent information on the stars (ST, \vsini, \bv) and on the data obtained as well as various measurements (see below).


\subsection{Measurements}

 \subsubsection{Radial velocities}
 The extraction of the radial velocities is fully described in \cite{galland05a}. Briefly, for each star, we build a first estimate of the reference spectrum which is the average of the spectra recorded and reduced via the STS {\small HARPS}  pipeline. We then compute a first estimate of the RV for each spectrum, by correlating in the Fourier space each spectrum and this first estimate of the reference spectrum. We then build a final reference spectrum by averaging the spectra once shifted from their measured RV. For each spectrum we finally measure the RV velocity with respect to this reference spectrum. We also measure the uncertainties associated to each RV measurement.

Note that to build up the reference we compute the $\chi^2$ of each spectrum compared to the first estimate of the reference spectrum. Most of the time, the $\chi^2$ found is much less than 10. Whenever a higher $\chi^2$ was found, we checked the spectra. In such cases, either they were due to bad observing conditions or technical problems and were not kept to build the reference spectrum (this actually happened quite rarely as we already selected spectra with acceptable absorptions) or they were associated to lines deformations indicative of a type-2 binary. 

\subsubsection{CCFs, bisectors and bisectors velocity-span}
Whenever possible (see below) 
we computed for each target the resulting cross-correlation functions (CCFs) and the bisectors velocity-span (see for their definition \cite{galland05a}). Indeed, the bisector and bisectors velocity-span are very good diagnostics of stellar activity (spots, pulsations) provided {\it 1)} they can be measured (see below), and {\it 2)} the star projected rotational velocity is larger than the instrumental resolution (see \cite{desort07}).

The uncertainty associated to the bisectors velocity-span depends directly on the projected rotational-velocity and/or their spectra type. Indeed, the number of lines used to compute the CCF depends on these two parameters (much more than on the signal-to-noise ratio). For stars with high \vsini~(typ. $\geq$ 150\,\kms) and/or \bv~$\leq$ 0.1, the number of lines may be quite low (30--50) whereas for late-type stars with moderate \vsini~(10--20\,\kms), the number of lines used is a few hundreds (up to about 1\,000). When the bisectors were computed, we then attributed quality flags to the bisectors velocity-span measurements, respectively: Good, Acceptable, Bad, corresponding to numbers of lines respectively $\geq$ 100, 40--100, and $\leq$ 40.

\addtocounter{table}{1}

\subsection{Diagnostics for the classification of variable stars}

Variable stars are defined as having a RV standard deviation (rms) larger than twice the RV uncertainties and a total RV amplitude larger than 6 times the RV uncertainties. RV variations can {\it a priori} be due to the presence of a companion (star, brown dwarf, planet) or to intrinsic variations of the star (spots, pulsations). It can also be a combination of those different origins.

\subsubsection{Binarity}
\begin{figure}
  \centering
  \includegraphics[width=\hsize]{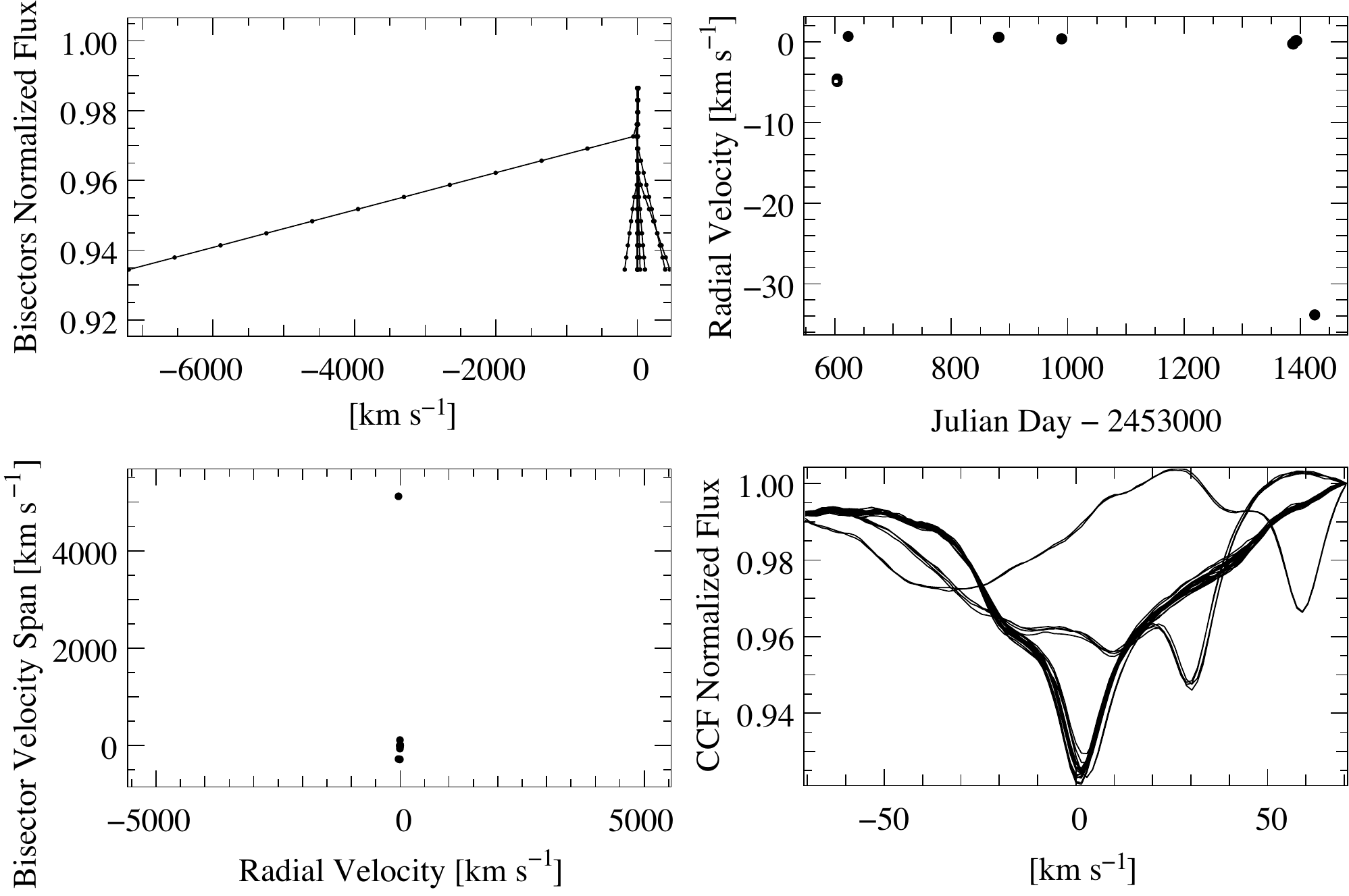}
  \caption{Example of an SB2 binary, HD\,2885 (A2V; \vsini~= 40\,\kms). RV curve (upper right), CCFs ({\bf lower right}), bisectors ({\bf upper left}), and bisectors velocity-span (lower left). The CCF is clearly variable and indicative of an SB2 binary.}
  \label{binary_chi2_hd2885}
\end{figure}

 We first checked those stars with high $\chi^2$ ($\geq$ 10) and looked for line deformations indicative of spectroscopic binaries. Figure~\ref{binary_chi2_hd2885} provides an example of a binary SB2 identified on the basis of the $\chi^2$, HD\,2885 (A2V; \vsini~= 40\,\kms). It has to be noted that in such cases, the RV values measured are not any longer valid, as our RV extraction method assumes that all lines in a given spectrum originate from the same object.

For the rest of the variable stars, we tried to identify binary stars among the stars for which the RV amplitude can be explained by the presence of a stellar or BD companion. To do so, using rough estimations of the star masses via their \bv, we computed the RV amplitude 2$\times$$K_{\rm 2d}$ expected from the presence of a 13\,\Mjup~body orbiting with a period of 2 days and the RV amplitude 2$\times$ $K_{\rm 200d}$ expected in the case of a 200-days period and compared these quantities to the observed RV amplitudes, once corrected from the RV variations observed within a night (in practice, over a few hours), as the variations occuring within a few hours are assumed to be due to stellar origin, see below. Quantitatively we define as ``in-night''~RV~amplitude for a given object, the amplitude of the nightly RV variations. We computed then the following quantities: $R_{\rm 2d}$ = (observed~RV~amplitude $-$ ``in-night''~RV~amplitude)/ 2 $\times$$K_{\rm 2d}$ and $R_{\rm 200d} $ = (observed~RV~amplitude $-$ ``in-night''~RV~amplitude)/ 2$\times K_{\rm 200d}$, to be used as thresholds to identify the binaries. Note that we chose 2 and 200-day periods as they are quite relevent given our temporal sampling and our average time baseline.

\begin{figure}
  \centering
  \includegraphics[width=\hsize]{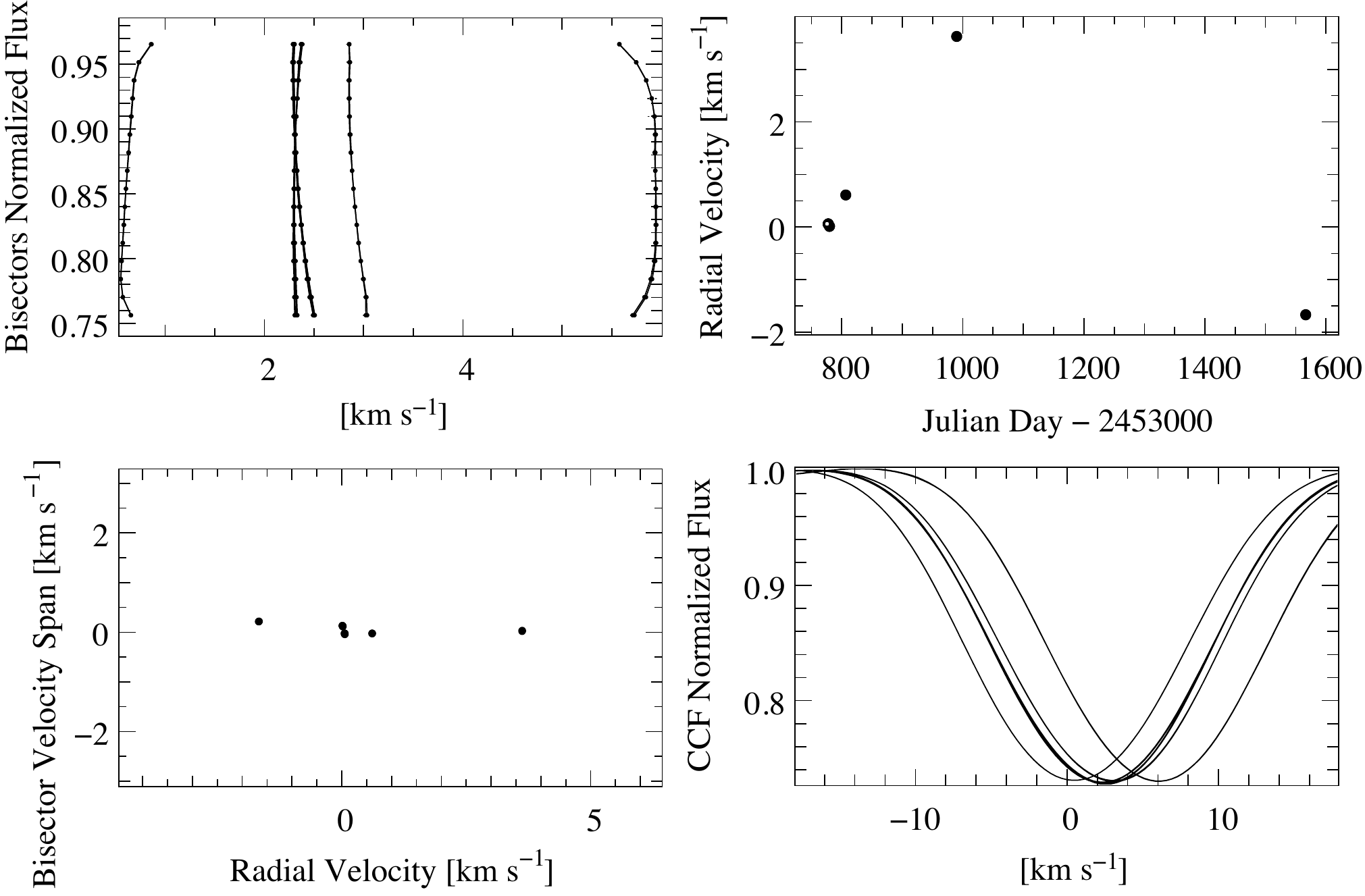}
  \caption{Example of a binary identified by a flat bisectors velocity-span diagram and high amplitude RV variations , HD\,68456 (F5V; \vsini~= 12\,\kms). RV curve (upper right), CCFs ({\bf lower right}), bisectors ({\bf upper left}), and bisectors velocity-span (lower left). The mass of the companion falls in the stellar domain (see text).}
  \label{binary_span_hd68456}
\end{figure}

For those variable stars for which we could compute a CCF and test the relation between bisectors velocity-span and RV variations, we selected those that show a simple, flat bisectors velocity-span, \ie values of bisectors velocity-span arranged horizontally in a (RV; bisectors velocity-span) diagram (this corresponds to stars for which the ratio of the amplitude of the bisector velocity span to the RV amplitude is smaller than 0.2) and for which $R_{\rm 2d}$ $\geq$ 2 or $R_{\rm 200d}$ $\geq$ 2; we regard them as unambiguous binaries. Figure~\ref{binary_span_hd68456} provides an example of a variable star (HD\,68456; F5V; \vsini~= 12\,\kms) for which the bisectors velocity-span clearly indicates the presence of a companion and the observed RV amplitude once corrected from in-night variations can be due to a $\simeq$ 0.1 \Msun~stellar companion (see below). Note that this star was also recently classified as a binary on the basis of astrometric data (see below; \cite{goldin07}).

\begin{figure}
  \centering
  \includegraphics[width=\hsize]{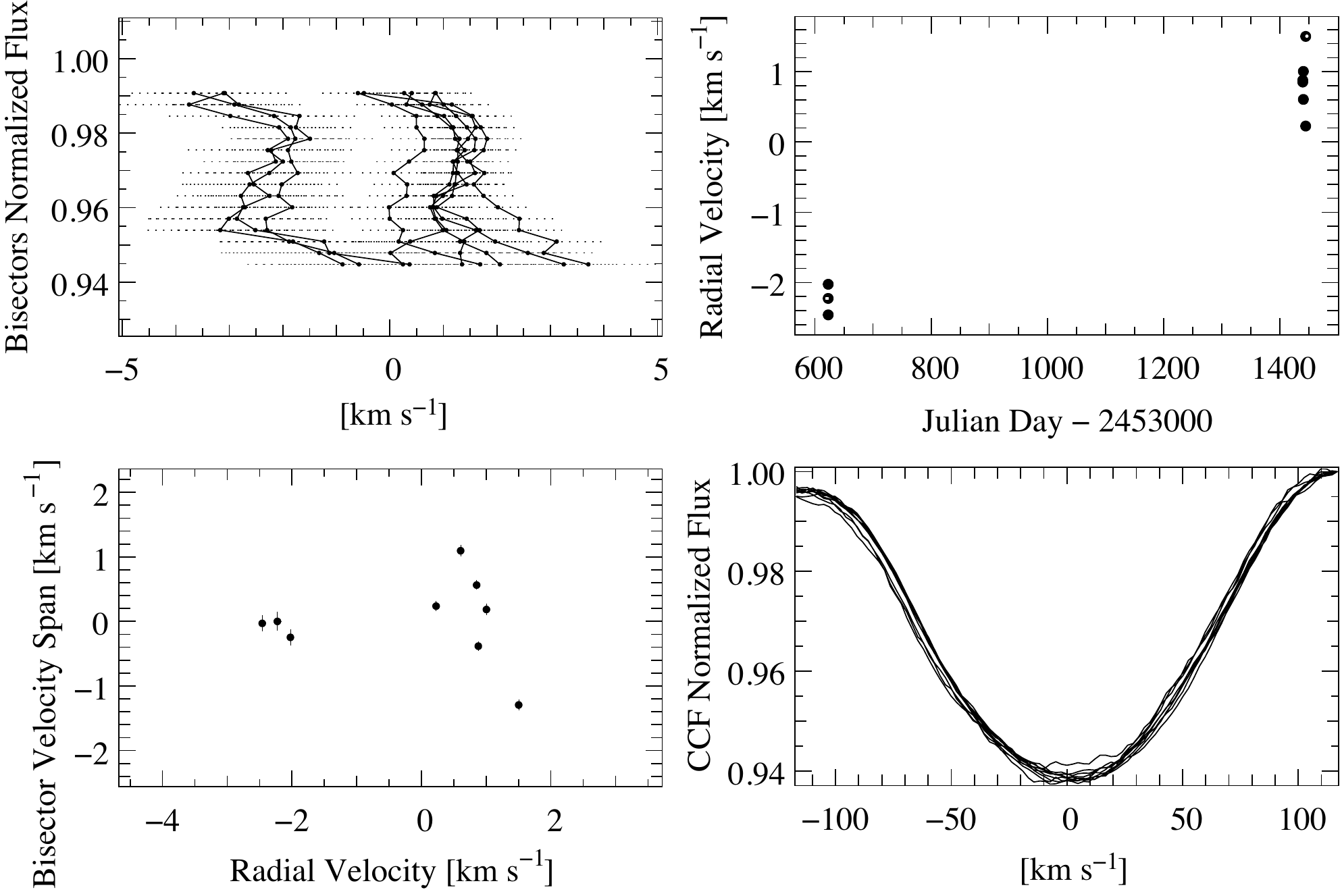}
  \caption{Example of a star whose RV variations are most probably due to both pulsations and binarity, HD\,19545 (A3V; \vsini~= 80\,\kms). RV curve (upper right), CCFs ({\bf lower right}), bisectors ({\bf upper left}), and bisectors velocity-span (lower left). The CCFs are clearly variable; the bisectors velocity-span diagram is composite: part of the data are spread horizontally over a large velocity range, and part are spread vertically, over a large range of span . The points that give the vertical bisectors velocity-span are those associated to the nightly high frequency RV variations; their bisectors are strongly variable in shape. The points with the low RV are associated to bisectors spans that are clearly shifted from the ones corresponding to higher velocities; this produces a shifted bisectors velocity-span.}
  \label{binary_puls_span_hd19545}
\end{figure}

\begin{figure}
  \centering
  \begin{tabular}{cc}
    \includegraphics[width=4cm]{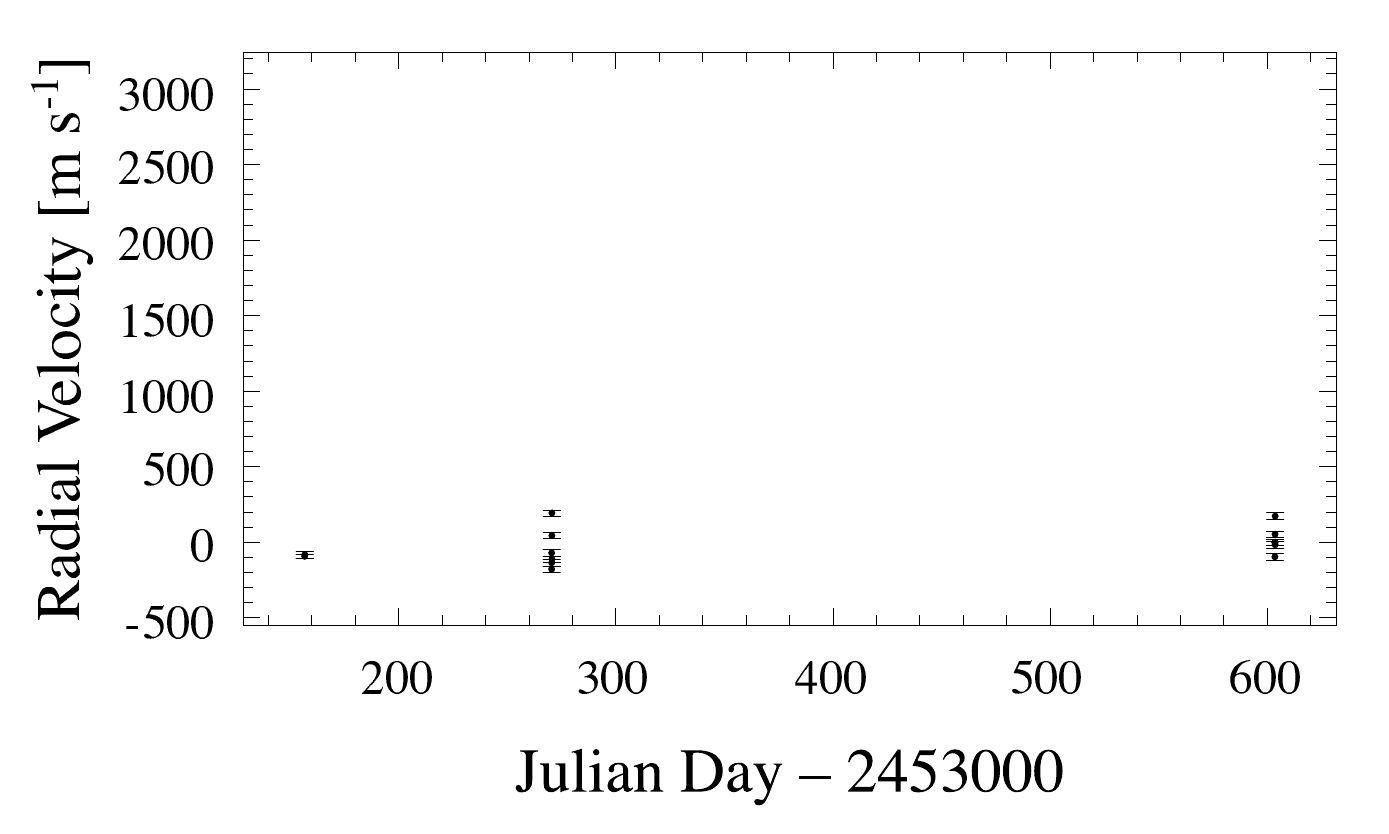} &
    \includegraphics[width=4cm]{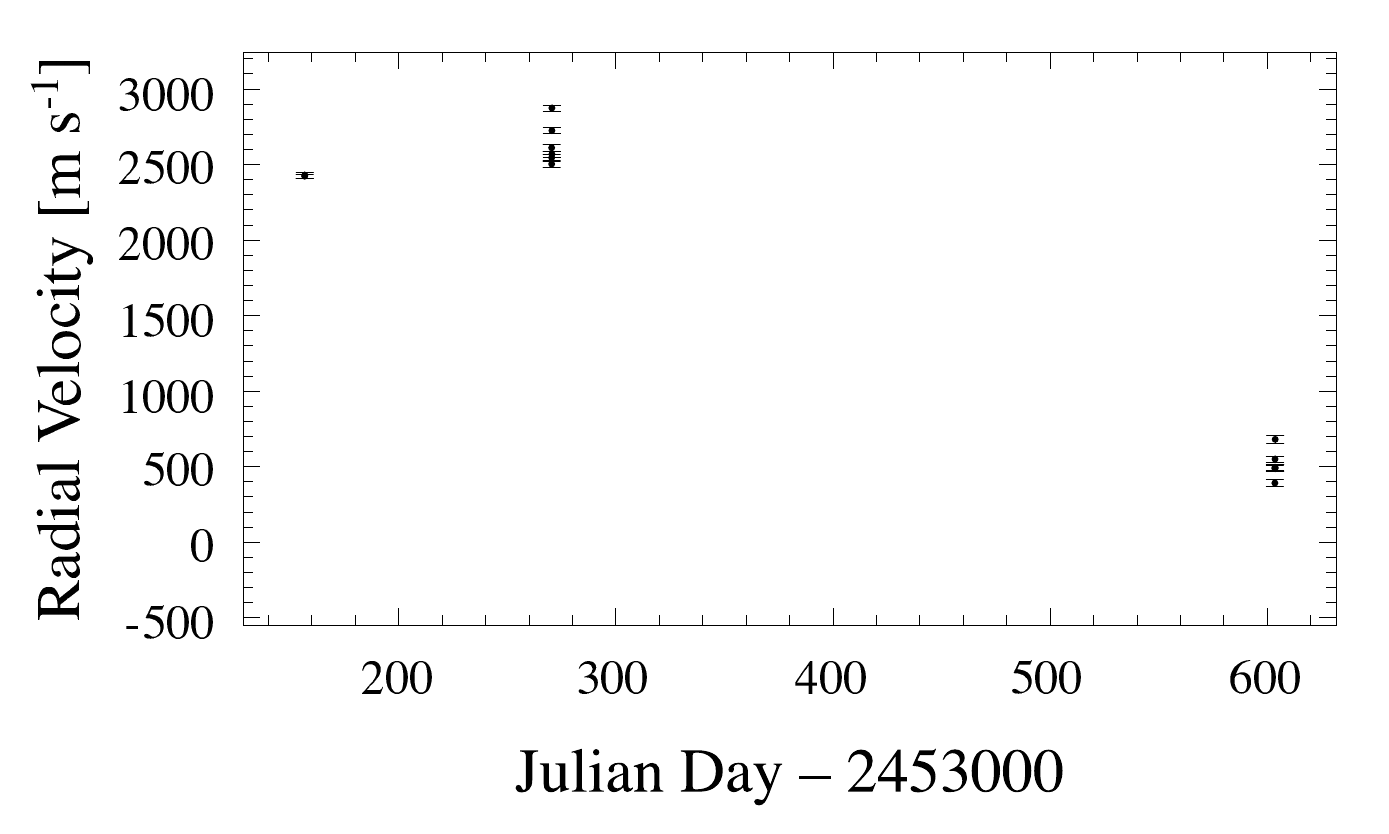} \\
    \includegraphics[width=4cm]{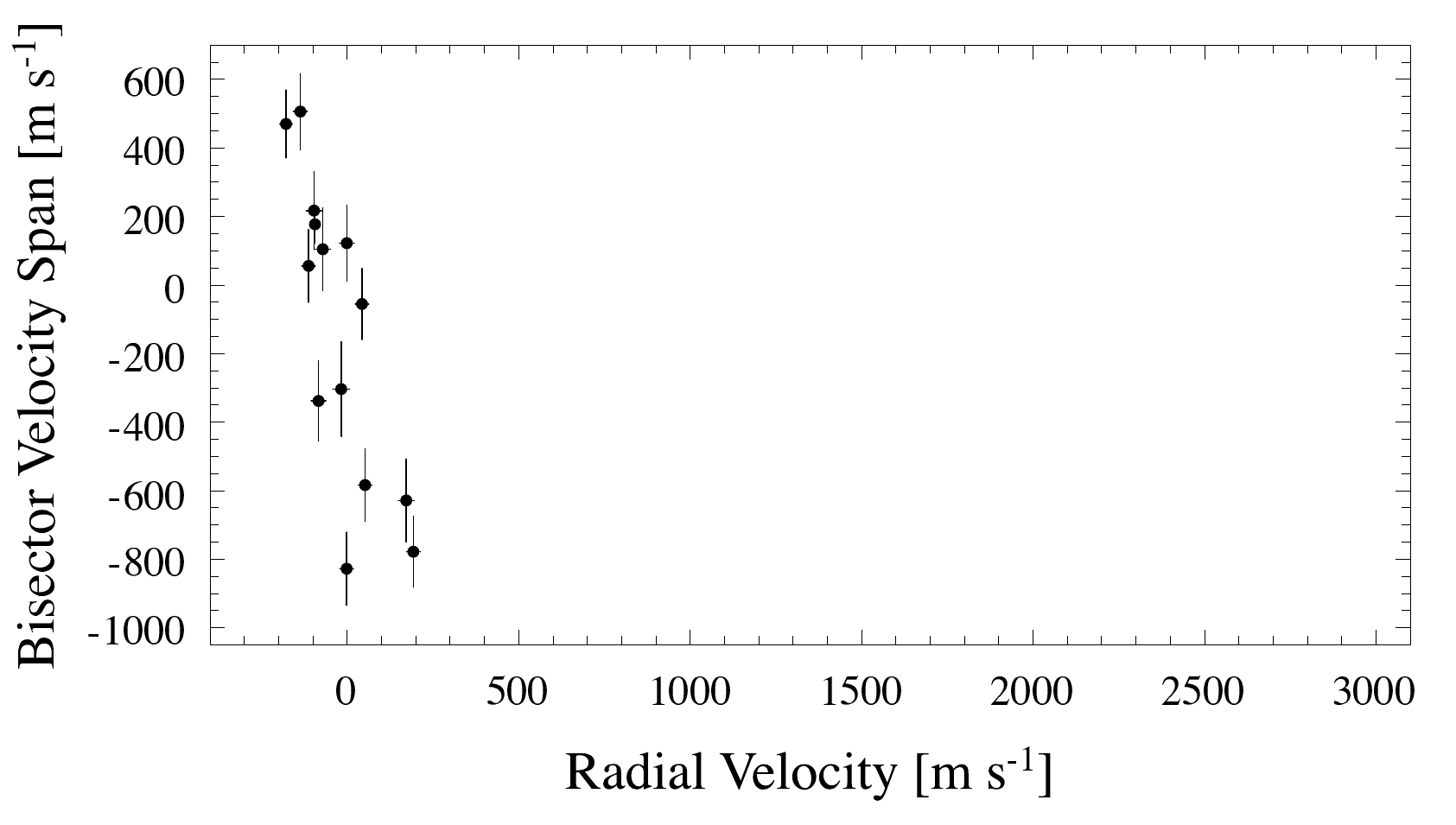} &
    \includegraphics[width=4cm]{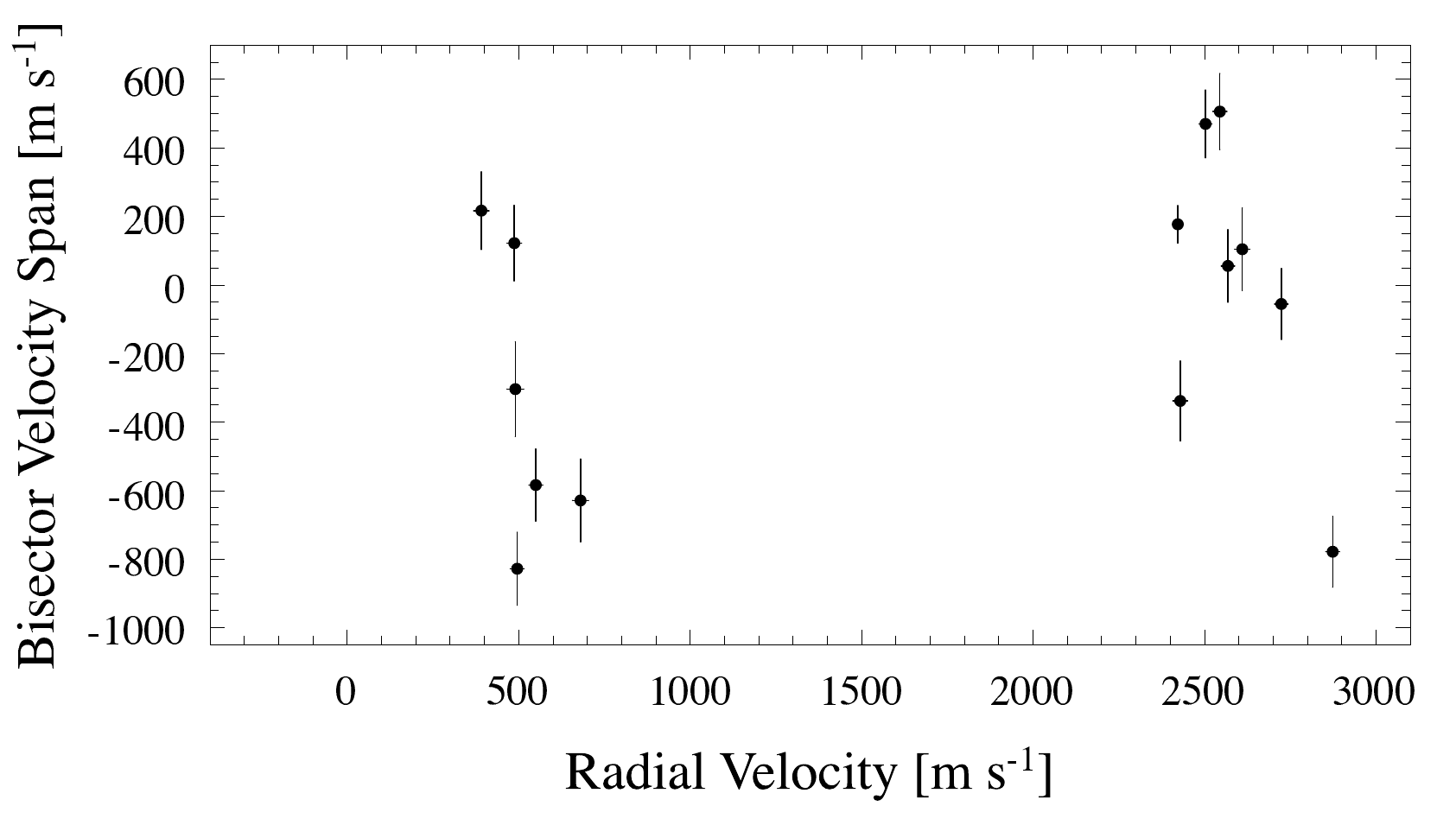}
  \end{tabular}
  \caption{Simulation of a composite bisectors velocity-span diagram produced when adding a 100-\Mjup~companion on a circular orbit, with a 120-day period around a 1.8-\Msun~pulsating star, HD\,159492 (A7V; \vsini~= 60\,\kms). The initial RV and bisectors velocity-span data are showed on the left; we see in particular high frequency (nightly) RV variations and bisector velocity-spans spread vertically; the simulated data are showed on the right. The bisectors velocity-span diagram on the right is clearly composite: both flat over a wide range of RV + vertical over a wide range of span values,  similarly to HD\,19545.}
  \label{composite_hd159492_orig_plan}
\end{figure}

Some variable stars show a bisectors velocity-span which is either partly flat and partly vertical or partly flat and partly inclined, indicating that most probably they are binaries and at the same time pulsating or active (see below). We classify those objects with $R_{\rm 2d}$ or $R_{\rm 200d}$ $\geq$ 2 as strong binaries candidates. An example, HD\,19545 (A3V; \vsini~= 80\,\kms), is provided in Fig.\,\ref{binary_puls_span_hd19545}. Figure~\ref{composite_hd159492_orig_plan} shows for comparison the case of a pulsating star for which we artificially simulated an additional companion star. The generated RV and span curves of the pulsating star and pulsating star plus stellar companion are comparable to those found in the case of HD\,19545. Note that we do not have quantitative criteria to identify those ``composite bisectors velocity spans''; this is why we classify the candidates as strong candidates rather than unambiguous binaries.

\begin{figure}
  \centering
  \includegraphics[width=0.6\hsize]{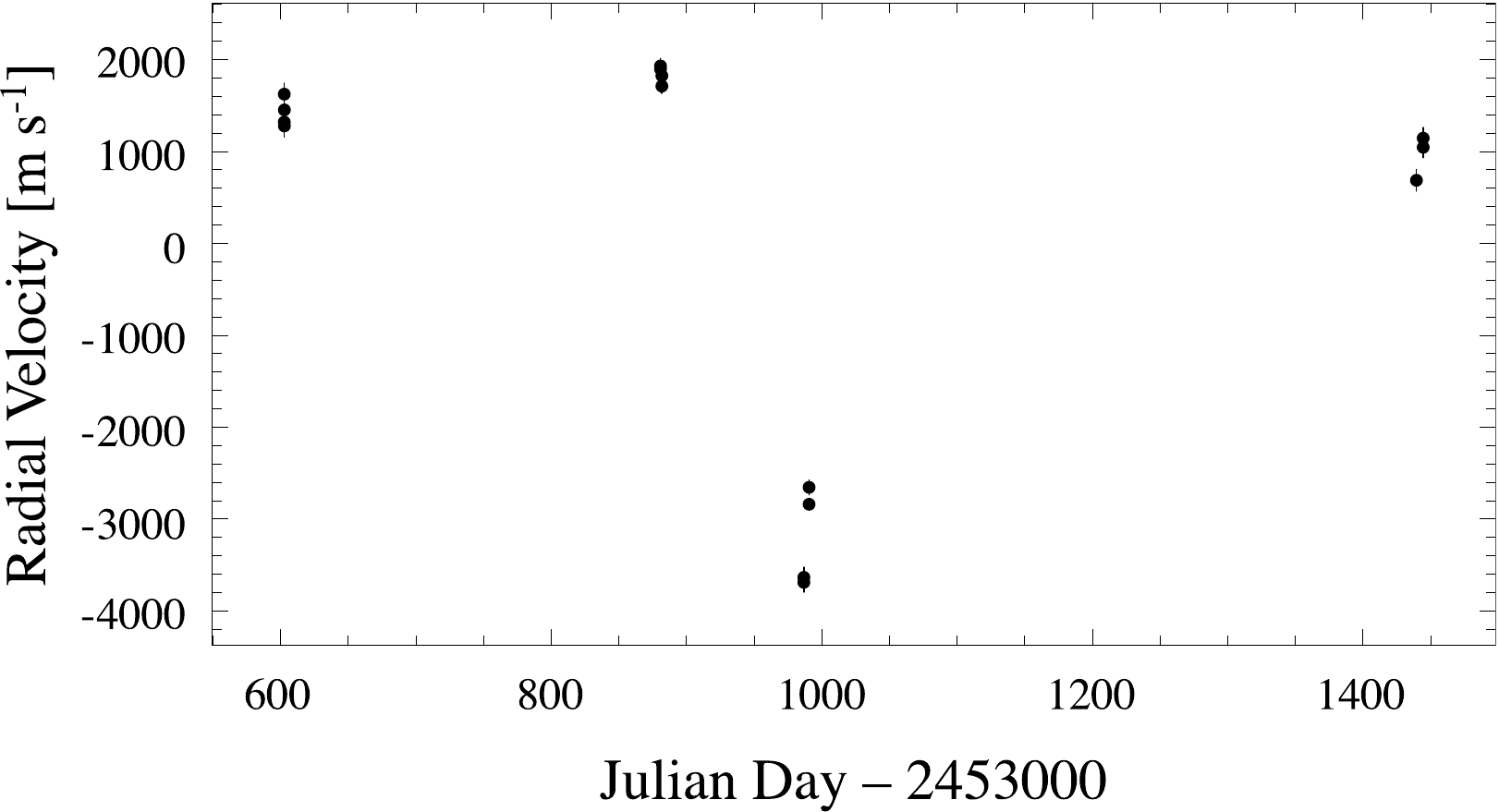}
  \caption{Example of a star whose RV variations are most probably due to binarity, HD\,200761 (A1V; \vsini~= 80\,\kms). RV curve. This is a case where no CCF could be computed, and the binarity classification relies solely upon the RV curve.}
 \label{binary_rv_hd200761}
\end{figure}

Finally, for the rest of the stars, we flagged as binary candidates those stars with $R_{\rm 2d}$ or $R_{\rm 200d}$ $\geq$ 4. Note that for these stars we conservatively adopted a more stringent threshold for $R_{\rm 2d}$ or $R_{\rm 200d}$ as we lack of additional indication of the presence of companions, and we know that these stars may be intrinsically variable; we thus took into account the fact that the actual amplitude RV variations due to the pulsations may be larger than the one measured on our set of data. This ensures that most of the observed RV amplitudes is due to a perturbation by a BD or a star. Figure~\ref{binary_rv_hd200761} gives an example of such a star.


\subsubsection{Planets}
Those stars that shows at the same time signs of RV variability with low amplitudes and a flat bisectors velocity-span diagram are very good candidates for hosting planets. Note that in some cases, stars showing composite bisectors velocity-spans diagram with $R_{\rm 2d}$ or $R_{\rm 200d}$ larger than 2 can still be ``good'' candidates for hosting planets. In such cases, the total RV amplitude is not dominated by the planetary signatures but but stellar variability (\eg spots).

\subsubsection{Intrinsically variable stars}
\begin{figure}
  \centering
  \includegraphics[width=\hsize]{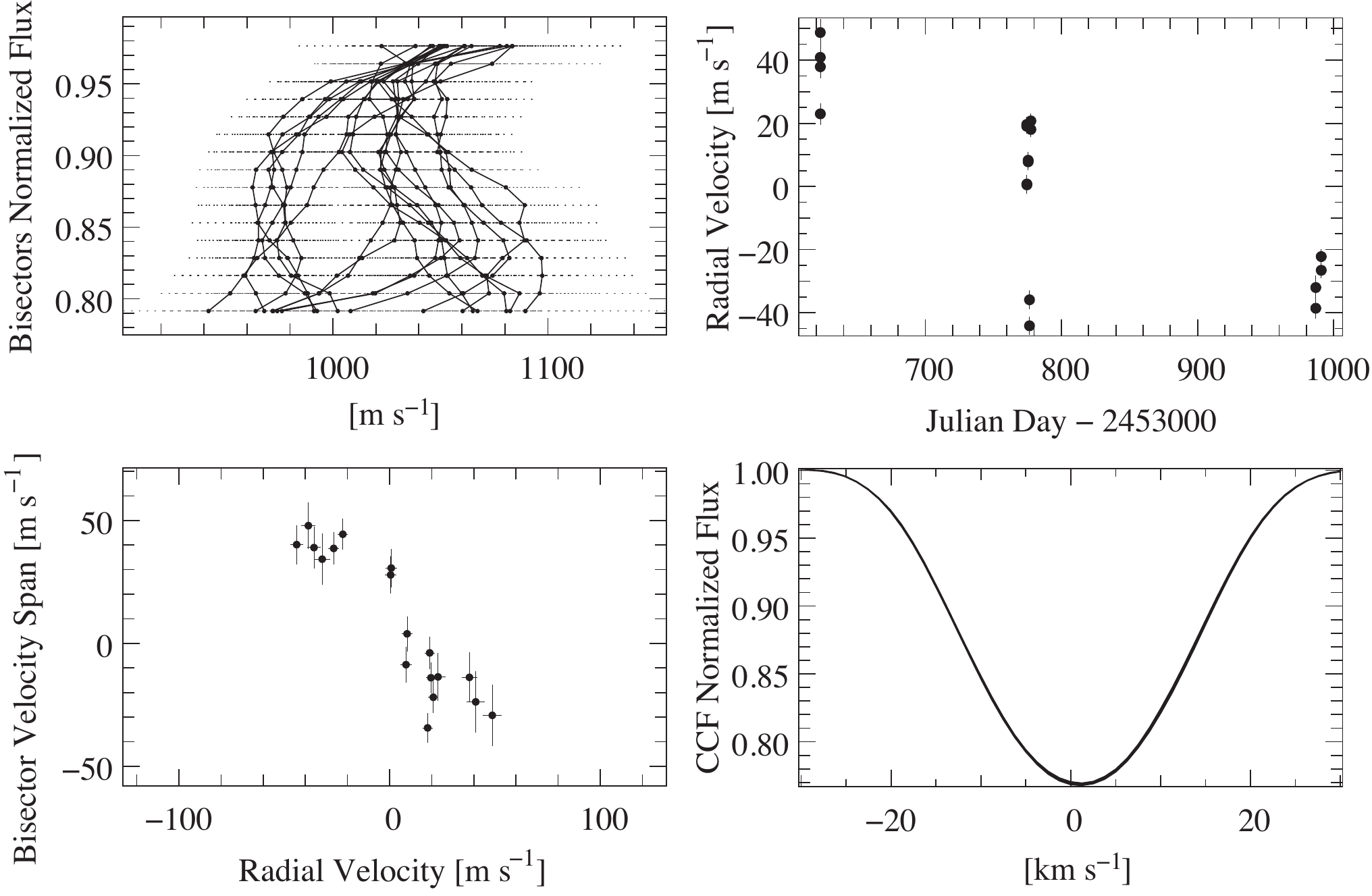}
  \caption{Example of a star with variable RV due to the presence of spots. HD\,25457 (F5V; \vsini~= 25\,\kms). RV curve (upper right), CCFs ({\bf lower right}), bisectors ({\bf upper left}), and bisectors velocity-span (lower left). The bisectors velocity-span variations are clearly correlated to the RV variations.}
  \label{spot_hd25457}
\end{figure}

In the case of spots, and provided the star \vsini~is higher than the instrumental resolution, the bisector shape is very peculiar and the bisectors velocity-span variations are correlated to the RV ones (see \cite{desort07}). In a (RV; bisectors velocity-span) diagram, the bisectors velocity-span values are arranged either as an inclined ``8'' shape, or along an inclined line (so called ``anti-correlation''). For these objects, the ratio of the bisector span amplitude to the RV amplitude is found to be in the range 1--3. Figure~\ref{spot_hd25457} provides an example of a star showing clear signatures of spots on the basis of the bisectors velocity-span diagram (HD\,25457; F5V; \vsini~= 25\,\kms). As another example, the very nice case of HD\,138763 can also be found in \cite{desort07}.

\begin{figure}
  \centering
  \includegraphics[width=\hsize]{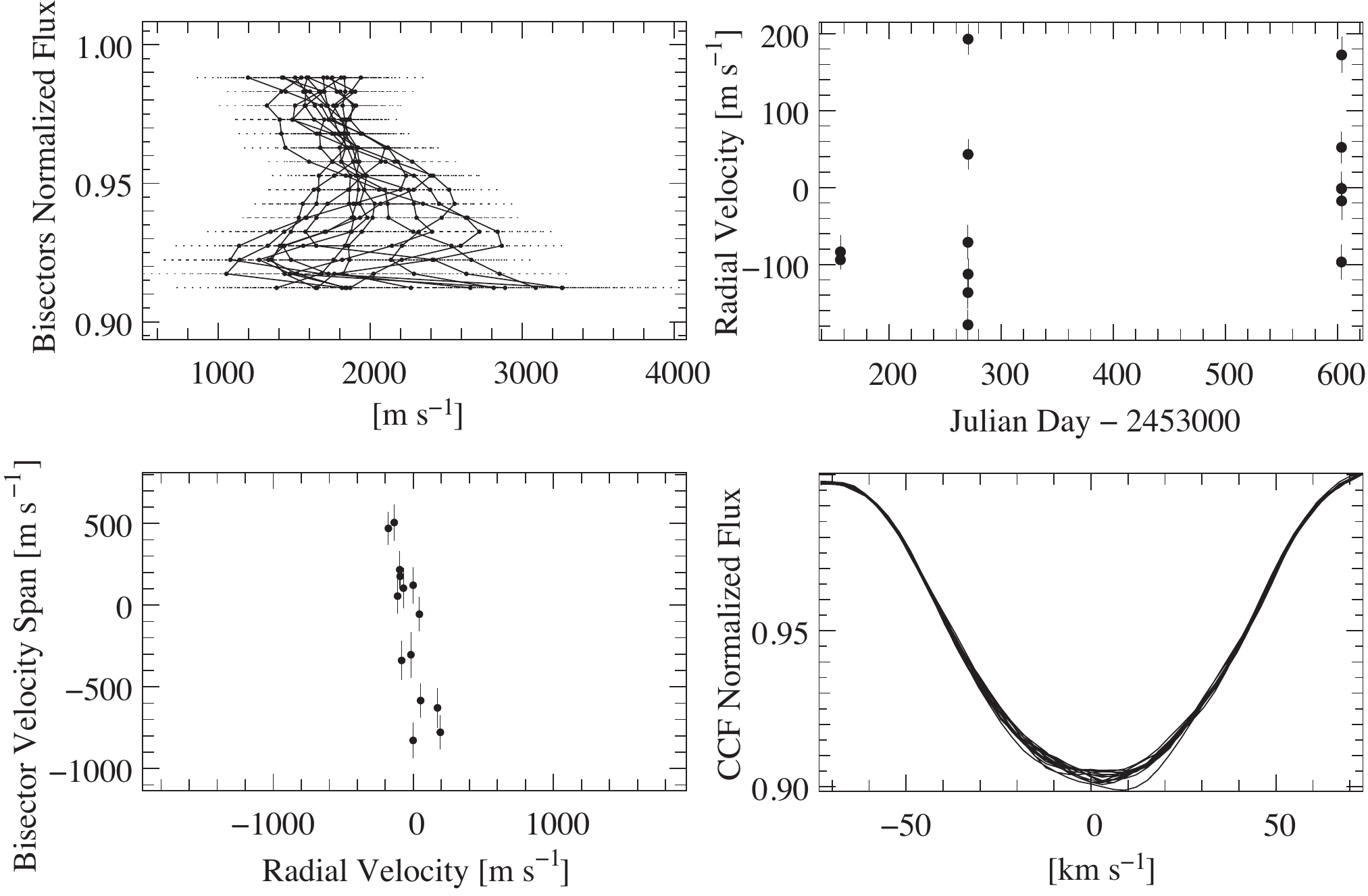}
  \caption{Example of a star with variable RV due to pulsations. HD\,159492 (A7V; \vsini~= 60\,\kms). RV curve (upper right), CCFs ({\bf lower right}), bisectors ({\bf upper left}), and bisectors velocity-span (lower left). The CCF and bisector variations are due to line deformations; the bisectors velocity-span variations are not correlated to the RV variations.}
  \label{puls_hd159492}
\end{figure}

In the case of pulsations, the bisectors velocity-span values are spread over a much larger range than the RV and their variations are not correlated to the RV ones. In a (RV; bisectors velocity-span) diagram, the bisectors velocity-span values are spread vertically, and the ratio of the bisector span amplitude to the RV amplitude is large, typically $\geq$ 3. Figure~\ref{puls_hd159492} provides an example of a pulsating star (HD\,159492; A7V; \vsini~= 60\,\kms).

\section{Results}
\label{results}
 Given the variability criteria described above, 108 stars out of 170 are found to be variable in RV, and 62 are found to be constant in RV within our precision limits. Table\,\ref{sample} provides relevent measurements on these targets: RV amplitudes and uncertainties, bisector velocity-span rms and uncertainties.

\subsection{Variability classification}
\subsubsection{Stellar binaries}
20 stars are identified as binaries or candidate binaries with the criteria given in the previous section. More precisely:
\begin{itemize}
\item[-] 4 binaries are found on the basis of the $\chi^2$ criterium, namely HD\,99453, HD\,209819, HD\,2885 (Fig.\,\ref{binary_chi2_hd2885}), HD\,142629. 
\item[-] 6 stars show mostly flat bisectors velocity-span in a (RV; bisectors velocity-span) diagram: HD\,11262, HD\,68456, HD\,41742, HD\,116568, HD\,216627, HD\,54834. Their RV amplitude varies between 1\,600 and 9\,200\,\ms.
\item[-] 4 stars have composite, flat+vertical bisectors velocity-spans in a (RV; bisectors velocity-span) diagram, together with a total RV amplitude dominated by the binarity effect. These pulsating binaries are: HD\,220729, HD\,12311, HD\,19545 (Fig.\,\ref{binary_puls_span_hd19545}) and HD\,112934.
\item[-] Finally, 6 stars are classified as probable binaries on the sole basis of their RV variations: HD\,158352, HD\,177756, HD\,158094, HD\,2834, HD\,200761 (Fig.\,\ref{binary_rv_hd200761}), HD\,116160.
\end{itemize}

\begin{figure*}[ht!]
  \centering
  \includegraphics[width=0.45\hsize]{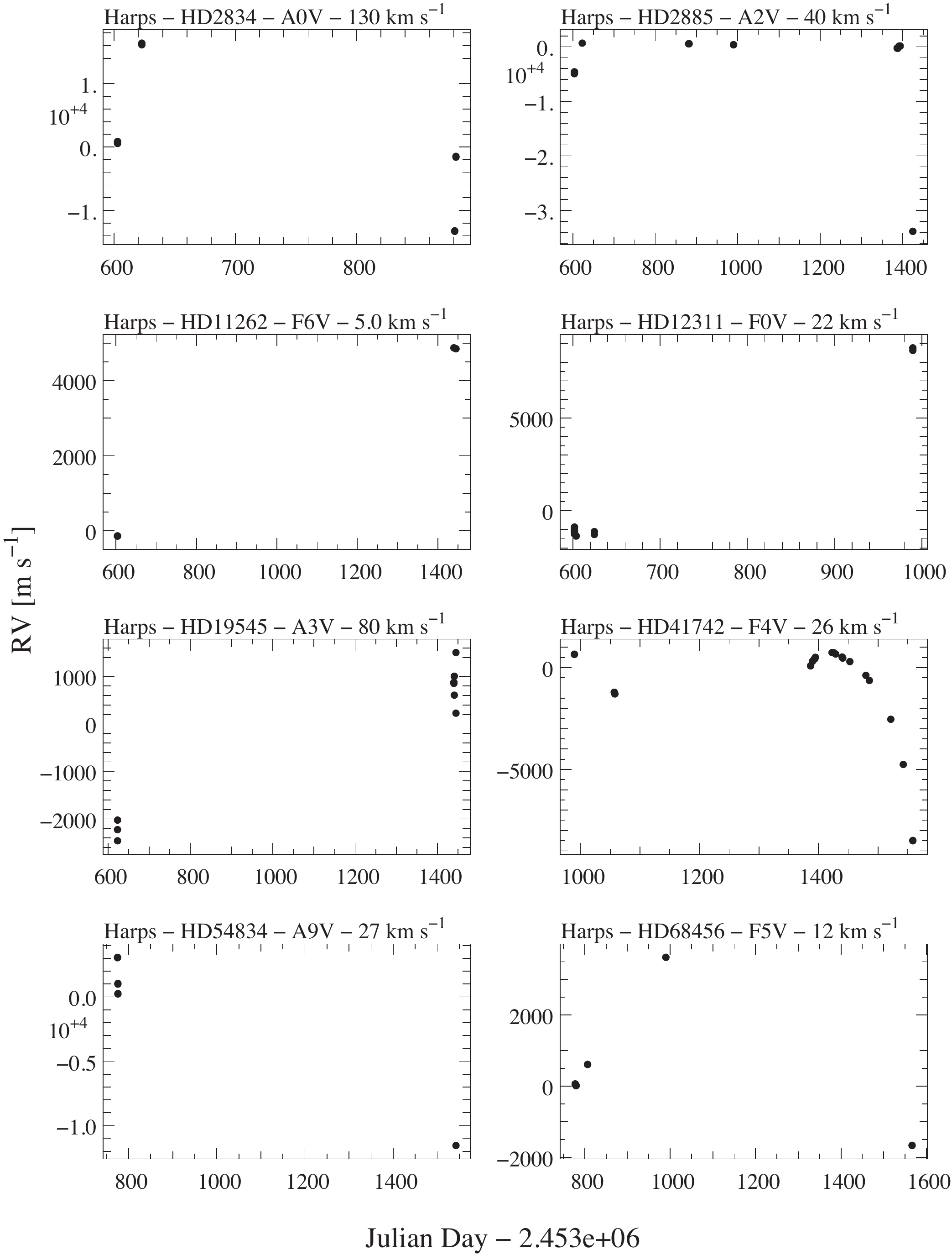}
\includegraphics[width=0.45\hsize]{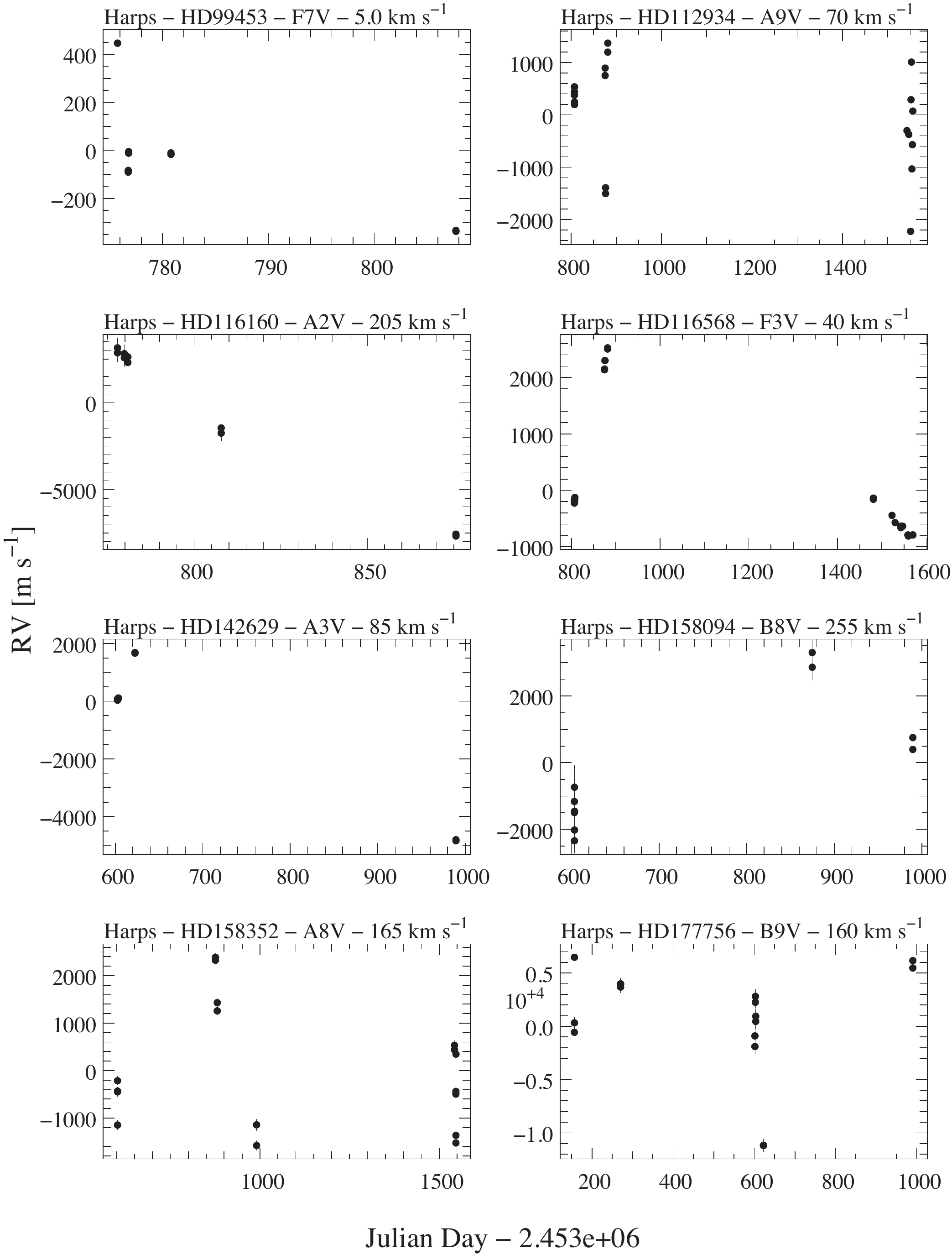}
\includegraphics[width=0.45\hsize]{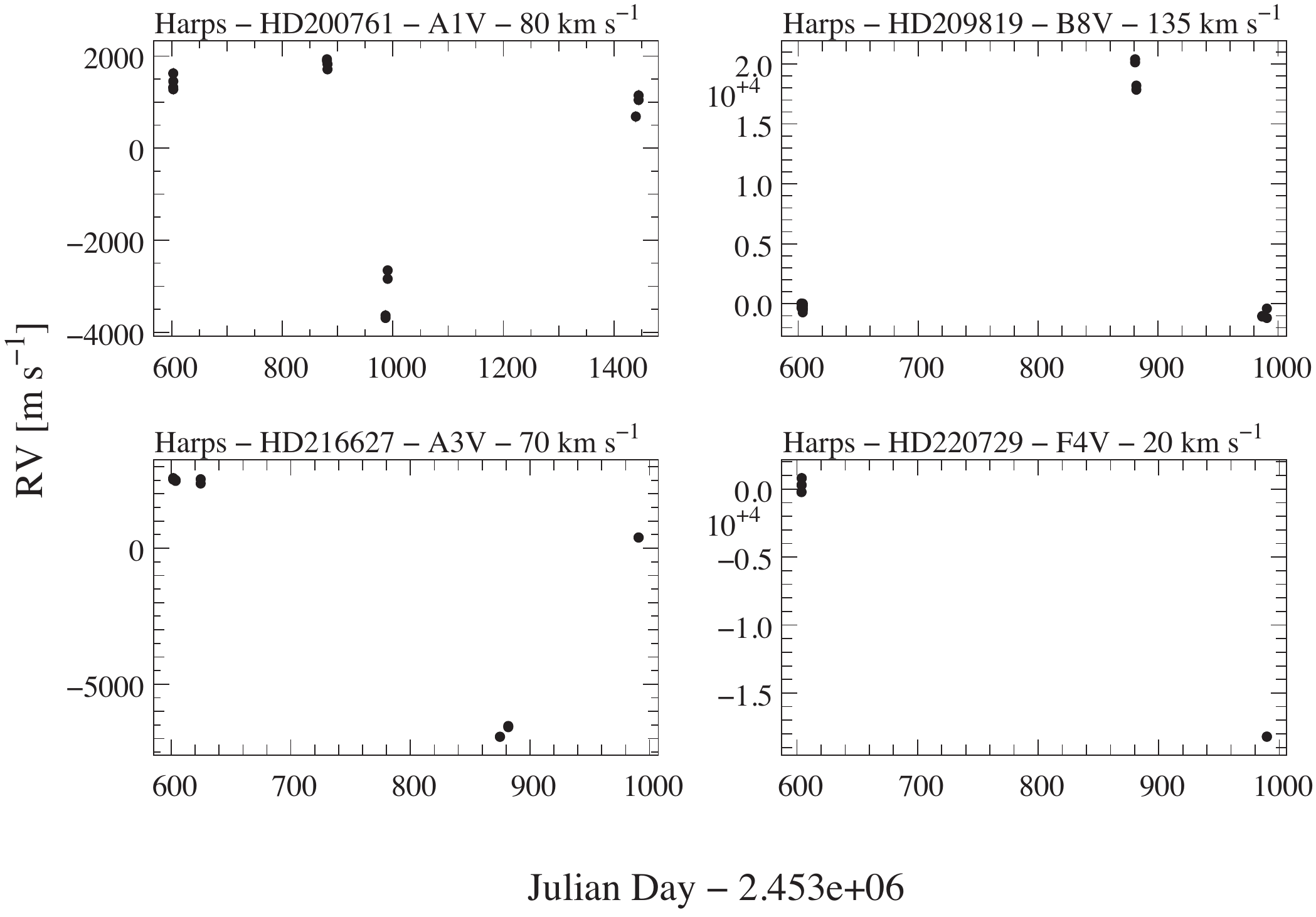}
  \caption{Radial velocity curves of the identified or strong candidates or probable binaries (see text).}
  \label{binaries}
\end{figure*}

The stars are flagged in Table\,\ref{sample}, and an indication on the criteria which was used to identify them as binaries or possible binaries is also given. No attempt was made to further characterize the stellar companion once the binary status was established, and no more data were recorded on the objects. Their RV variations are given in Figure~\ref{binaries}.

Notes on some individual binaries (and potential binaries):
\begin{itemize}
\item[-] HD\,11262 is associated to a ROSAT source by \cite{suchkov03}.

\item[-] HD\,54834: \cite{koen02} (2002) reported this star as a photometric Hipparcos variable at a level of 0.0046\,mag and with a frequency of 0.802\,day$^{\rm -1}$. Our data do not permit to confirm or infirm this frequency (not enough points, sampling not adapted).
\item[-] HD\,68456 (Fig.\,\ref{binary_span_hd68456}) was not reported as binary in the Hipparcos catalog from the photometric and astrometric points of view; it is classified by \cite{adelman01} as one of the Hipparcos least variable stars. \cite{goldin07} however provide an orbital solution to fit the Hipparcos astrometric data. The period found is 483$\pm$20\,days, a$_{0}$ = 9.6$^{+2.6}_{-1.2}$\,mas, eccentricity = 0.12$^{+0.25}_{-0.15}$, inclination = 131$\pm 16$$\degr$, $\omega$ = 103$^{+72}_{-68}$$\degr$ and $\Omega$ = 171$^{+164}_{-83}$$\degr$. Fixing the period and eccentricity proposed by these authors, we tried to find a fit to our RV data. They happen to provide good fits assuming a mass of $\simeq$ 100\Mjup~for the companion.
\item[-] HD\,99453: \cite{baade97} questioned the previously suggested SB2 status of this object on the basis of their data; we do confirm the SB2 status for this star.
\item[-] HD\,112934 (A9V; \vsini~= 70\,\kms): using Hipparcos photometry, \cite{handler99} reports this star as a new possible $\gamma$\,Doradus candidate but with a ``weak complicated signal'', associated to a 0.8-day period. \cite{decat06} did not find clear line-profile variations in their {\small CORALIE} data. From our data, the star is both pulsating and member of a binary system, which makes the line-profile variations indeed more complicated than for pulsating stars. Our limited number of data does not permit to characterize the high-frequency period.
\item[-] HD\,116160 was reported as an astrometric binary with accelerating proper motion by \cite{makarov05}.
\item[-] HD\,116568 was classified as one of the least variable stars with Hipparcos by \cite{adelman01}. \cite{baade97} report no variations in their $\pm$0.5\,\kms~spectroscopic survey. The present data show that this star is a binary with an amplitude of at least 2\,750\,\ms. It is also reported as an unresolved Hipparcos problem star by \cite{masson99} and associated to a ROSAT source by \cite{suchkov03}.
\item[-] HD\,142629 is an astrometric Hipparcos binary. It was also recently reported for the first time as a spectroscopic binary by \cite{antonello06}.
\item[-] HD\,158352 was classified as a possible Herbig AeBe star by \cite{the94}. \cite{corporon99} in a survey of RV variations among Herbig AeBe stars did not find variations to a 5--10\,\kms~level. This star was reported as being surrounded by a dusty disk by \cite{oudmaijer92}, and \cite{moor06} give an age of 750$\pm$150\,Myrs for the system.
\item[-] HD\,177756 was classified as a possible $\lambda$\,Bootis star, as well as a possible SB (\cite{farragiana04}, \cite{gerbaldi03}). It is reported as one of the Hipparcos least variable stars (\cite{adelman01}).
\item[-] HD\,200761 was reported as one of the Hipparcos least variable stars (\cite{adelman01}).
\item[-] HD\,209819 was also reported as one of the Hipparcos least variable stars (\cite{adelman01}).
\item[-] HD\,220729 is associated to a ROSAT source by \cite{suchkov03}.
\end{itemize}

\subsubsection{Stars with planets}
One star, HD60532 (F6IV--V; \bv~= 0.52) clearly reveals at the same time low-amplitude RV variations and flat bisectors velocity-spans diagram, indicative of the presence of two Jupiter mass companions with a high-confidence level. This star and the results of the fits of the RV curve is presented in \cite{desort08a}. Interestingly in the frame of the present paper, the periods of the detected planets are long ($\geq$ 100\,days). 
Hence, we get at least 1\,$\%$ of F stars with long-period planets in our sample. This is much smaller than the predicted rate of $\simeq 10\,\%$ for 1.5\,\Msun~stars by \cite{kennedy08}; however, we are yet not sensitive to all range of masses and periods as will be showed in the last section.

\subsubsection{Single stars: intrinsic variability}
\begin{figure}[ht]
  \centering
 \includegraphics[width=.8\hsize]{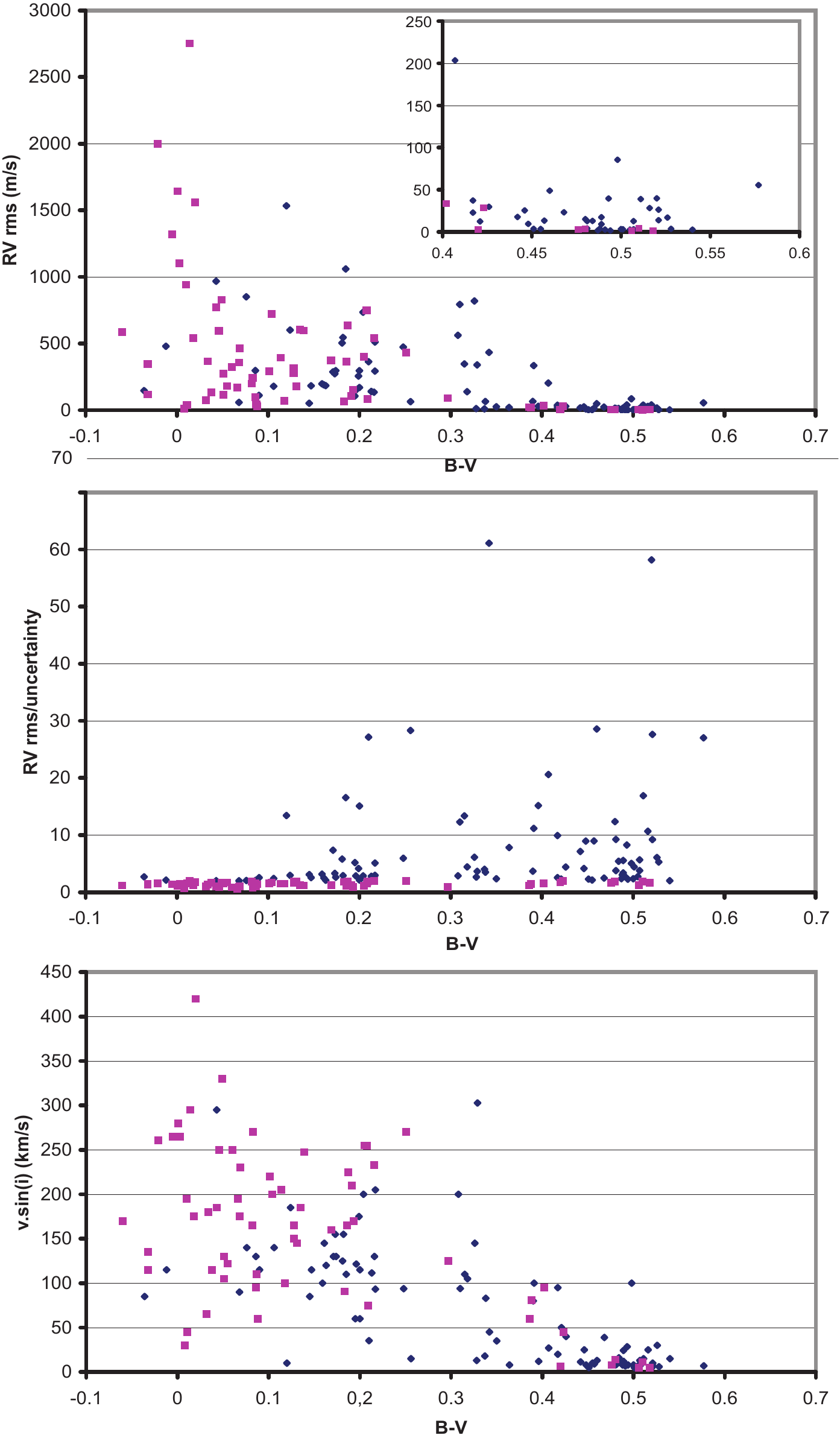}
\caption{Top: RV rms measured for all stars but binaries with more than 6 spectra available as a function of \bv. Middle: Ratio RV rms/uncertainty for the same stars. Bottom: (\bv; \vsini) diagram for the same stars. Losanges indicate RV variable stars and square indicate RV constant stars.}
  \label{fig_rms_BmV_allbutbin}
\end{figure}

We report in Table\,\ref{sample} the RV rms values obtained for each star, together with the associated uncertainties.
Figure~\ref{fig_rms_BmV_allbutbin} provides for all the stars except those identified as binaries the measured RV rms as a function of their \bv, the ratio RV rms/uncertainties (E/I) as a function of their \bv~as well, and the (\bv; \vsini) diagram for the same objects. In the plots we have distinguished the 88 stars that are found to be variable according to the criteria defined above and those 62 found to be constant according to the same criteria. 

The E/I ratio varies between 1.5 and a few tens; it is relatively smaller for stars with small \bv~than for those with larger \bv. More quantitatively, the median value for this ratio computed on variable stars is 2.7 (resp. 4.4 and 5.2) for stars with \bv~$\leq$ 0.2 (resp. 0.2 $\leq$ \bv $\leq$ 0.4 and \bv $\geq$ 0.4). Hence we detect more variable stars among stars with large \bv~than stars with smaller \bv. We see moreover that the uncertainties generally increase with decreasing \bv. 
 These results are not surprising and illustrate the fact that it is more difficult to identify variable stars when they have large uncertainties. In the frame of this study, it is important to keep in mind that our ability to detect variability generally decreases with decreasing \bv. 

Note also that the uncertainties increase with increasing \vsini; we could actually verify that the uncertainties vary as \vsini~with a $(v\sin{i})^\alpha$ law where $\alpha = 1.5 \pm 0.1$, as predicted in \cite{galland05a}.


The percentage of variable stars depends on B$-$V in the following way:
\begin{itemize}
\item[-] Most (85\,$\%$) of the 58 stars with \bv~larger than 0.4 are found to be variable and the RV uncertainty is 2\,\ms (median value). Evenmore, 90\,$\%$ of the 46 stars with \bv~larger than 0.45, \ie well beyond the instability strip, are found to be variable and their uncertainty is 1.4\,\ms (median value). We conclude then that at a level of precision of 2\,\ms or less, most of the stars with \bv~larger than 0.4 are RV variable.
\item[-] Among the stars with \bv~between 0.2 and 0.4, the number of variable stars found is small, but this is due to a selection effect as known $\delta$\,Scuti and $\gamma$\,Doradus stars were removed from our sample (see above).
\item[-] Only 36\,$\%$ of the 73 stars with \bv~smaller than 0.2 are found to be variable. The percentage of variable decreases to 20\,$\%$ if we consider the 40 stars with \bv~smaller than 0.1. For those stars with \bv~between 0.1 and 0.2, we get as many variable as constant stars. The number of stars found to be constant according to our criteria increases then with decreasing \bv. However, we have seen that our ability to detect variable stars decreases with decreasing \bv. More quantitatively, the median uncertainty in the case of ``constant'' stars is $\simeq$ 290\,\ms~whereas the median uncertainty in the case of stars found to be variable is $\simeq$ 80\,\ms. Furthermore, the median uncertainty of constant stars is comparable to the median value of the standard deviation of variable stars (265\,\ms). Hence we may stipulate that in fact most of the stars with \bv~$\leq$ 0.2 are probably RV variable.
\end{itemize}

When the CCF and bisectors velocity-span criteria apply (in fact, whenever the bisectors velocity-span can be measured with a good or acceptable quality), we may try to further characterize the stellar variability. We find that in such cases, as expected, most of the variable stars with \bv~smaller than 0.3 show signs of pulsations, whereas most of the variable stars with higher \bv~show signs of spots.

\subsection{Variability of stellar origin and impact on planet detectability}
\begin{table*}[htp]
  \caption{ 
Median values of RV rms (second row) and RV uncertainties (third row) for stars with more than 6 spectra (3 epochs) available. The number of stars per bin of \bv~is also given (first row). Rows 7--9 give the median per bin of \bv~of the {\it achievable} detection limits deduced from the measured rms for each star, expressed in Jupiter mass. Three periods are considered: 3, 10 and 100 days (see Sect.\,\ref{results}). Note that we assume that the planet, supposedly on a circular orbit, is detectable if the amplitude of the RV variations is larger than $3\times RV rms $. Rows 7--9 give the percentage of stars for which the detection limit, given the measured RV rms, fall in the planetary domain.}
  \label{limits}
   \begin{center}
    \begin{tabular}{l l l l l l l l}
      \hline
      \bv		             & [$-$0.1; 0]& [0; 0.1] & [0.1; 0.2]  & [0.2; 0.3] & [0.3; 0.4] & [0.4; 0.5] & [0.5; 0.6]\\
      \hline
      number of stars	             & 7       & 31      & 32         & 16        & 17        &30        & 17 \\	
      median RV rms (\ms)            & 480     & 298     & 283        & 330       & 66        & 13       & 4\\	
      median RV uncertainty	(\ms)& 239     & 300     & 90         & 80        & 19        &2         &1.4 \\	
      median detection limit ($P$ = 3\,days)           & 10      & 5       & 4.5        & 5         &0.8        &0.17      &0.05 \\
      median detection limit ($P$ = 10\,days)         & 15      & 8       &7           & 7         &1.3        &0.25      &0.08\\
      median detection limit ($P$ = 100\,days)         & 31      & 17      &15          & 16        &2.8        &0.55      &0.17\\
      percentage (3\,days)           & 71      & 71      & 94         & 100       &100        &100       & 100\\     
      percentage (10\,days)          & 43      & 68      & 78         & 88        &88         &100       &100\\
      percentage (100\,days)         & 28      & 42      & 44         & 37        &60         &100       &100\\
      \hline
    \end{tabular}
  \end{center}
\end{table*}

The ``uncorrected'' jitters, as given directly by the measured RV rms are provided in Table\,\ref{sample}, for each star, together with the associated uncertainties. Note that we prefer not to use the jitters corrected from the uncertainties, as sometimes done, as our main aim is to evaluate the impact on planet detectabiliity rather than to make stellar studies.
Table\,\ref{limits} gives the computed median ``uncorrected'' jitters per bins of \bv. 

In \cite{galland05a} we had shown that the detection limit strongly depends on the star ST and its projected rotational-velocity; more precisely, the detection limit increases with earlier ST and/or larger \vsini. Thanks to the present data, we can in addition address the question of the impact of the stellar jitter.

\begin{figure}[ht]
  \centering
 \includegraphics[width=\hsize]{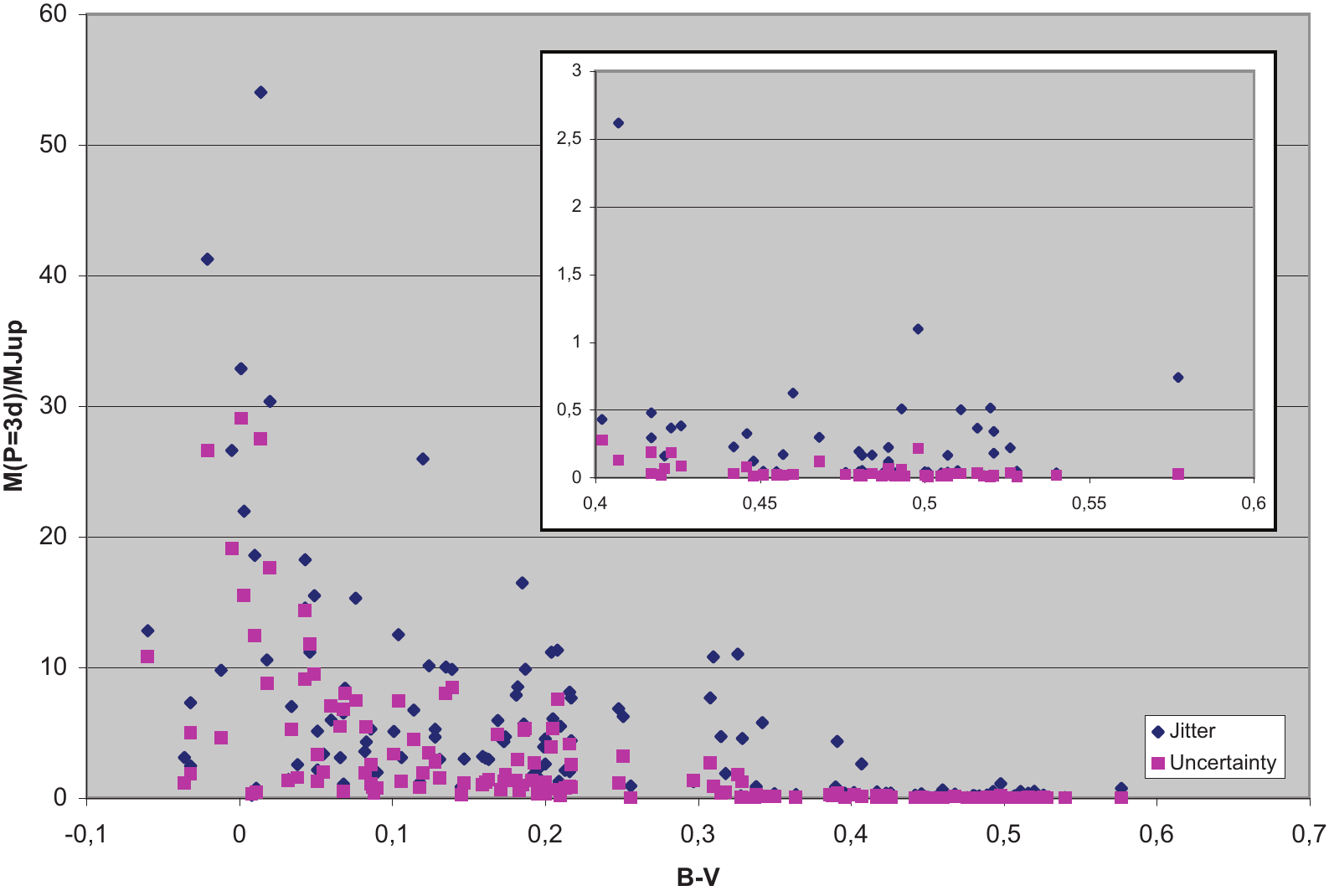}
\caption{Achievable detection limits for all stars but binaries taking into account their actual "uncorrected" jitter (losanges) or the measured uncertainties (squares), and assuming a planet on a circular 3-day orbit.} 
  \label{fig_limdet_BmV_3d_jit_uncert_allbutbin}
\end{figure}

To estimate the detectable masses for a given star and a given period, we assume that a planet, supposedly on a circular orbit is detectable if the amplitude (2$\times K$) of RV variations that it would produce is larger than 3$\times$ RV rms, where RV rms is the "uncorrected" jitter actually measured. We will come back later on the validation on this assumption. We give for each star in Table~3
the computed detectable limits assuming 3-day, as well as 10-day and 100-day periods. Figure~\ref{fig_limdet_BmV_3d_jit_uncert_allbutbin} shows the detection limits for all stars for a 3-day period. For comparison, we also give in this figure the mass of the planet that would be detectable if the star is not active/pulsating (hence has no jitter) and the limit would then be set by the uncertainty (hypothesis 2$K$ = 3$\times$ uncertainty). The ratio of the two values is E/I. As previously seen, this ratio is larger than 1.5 and may be quite high; the impact of the jitter on the detectable masses is therefore non negligeable.

Several comments can be made:
\begin{itemize} 
\item[-] the achievable 
limits fall into the planetary domain for a large number of stars: more precisely, in 137 out of 150 stars, \ie 91\,$\%$, the detection limit for a 3-day period falls within the planetary domain. For the remaining stars the limit falls well into the BD domain with masses up to 54\,\Mjup. When considering a 10-day (resp. 100-day) period, we find that we can reach the planetary mass domain for 124 stars, hence 83\,$\%$ (resp. 92 stars, hence 61\,$\%$). For a 10-day period, the limit for all remaining stars but one fall into the BD regime; for the 100-day period, the limit for all remaining stars but 6 fall in the BD domain. 
\item[-] as expected, the median of the detection limits {\it generally} improve with increasing \bv, from 10\,\Mjup~for \bv~between -0.1 and 0, to 5\,\Mjup~for \bv~between 0 and 0.3, to 0.05\,\Mjup~for \bv~between 0.5 and 0.6 (for a 3 day period) (see Table\,\ref{limits}). Noticeably, for stars with \bv~$\geq$ 0.3, individual detection limits may be as low as 0.02\,\Mjup and for stars with \bv~$\leq$, 0.3, individual detection limits may be as low as 0.5\,\Mjup. For a ten day period, these numbers become respectively: 15, 7 and 0.08\,\Mjup; for a 100 day period, 31, 16 and 0.17\,\Mjup. Also, noticeably, the detection limits steeply improves at \bv~= 0.3.
\item[-] the "uncorrected" jitter varies a lot from one object to the other. Therefore the general conclusion that the detection limits improves with increasing \bv~may not apply when considering individual objects: for instance, the two stars HD\,50445 (A3V; \bv~= 0.18) and HD\,63847 (A9V; \bv~= 0.3) have similar projected rotational-velocities (\vsini~$\simeq$ 90\,\kms) and very different levels of activity, with an RV rms of 66\,\ms~and 794\,\ms~respectively. When we take this "uncorrected" jitter into account, the detection limit is 1\,\Mjup~($P = 3$\,days) and 1.5\,\Mjup~($P = 10$\,days) around the A3V star whereas the detection limit is about 10 times higher for the A9V star.
\end{itemize}

We conclude then that planets can indeed be found around a wide range of stars with \bv~larger than $-$0.1, even taking into account their jitter. The achievable detection limit of such early type stars cannot be predicted given only the star properties (ST, \vsini), but requires to record data to estimate their level of jitter.

Note: we note that of course, the measured "uncorrected" jitter provides a reliable limit to planet detection only when this jitter is due to stellar activity in general, and not to companions. Would a companion be present, its contribution to the RV variability would have to be removed in order to estimate the impact of the stellar activity.


\section{Planet detection limits of the present survey}
\label{detectionlim}
\subsection{Estimation of the achieved detection limits}
We try now to estimate the detection limits reached by the present survey, taking into account the actual RV curve. For each star, we then compute the detection limit (companion mass) as a function of its period. To do so, we consider a planet with a given mass and with a given period (the orbit is assumed to be circular). For any couple (Mass; Period) we generate a large number of keplerian orbits, assuming different times of passage at periastron ($T_0$). For each orbit, we compute the expected radial velocities at the times of the actual observations. We add a noise (random value between $^{+}_{-}$ RV uncertainty), where RV uncertainty is the uncertainty measured on the RV data. We then get a virtual set of RVs, which takes into account the star properties (in particular, its ST and rotational velocity, through the uncertainties and SN). We compute then the standard deviation of the virtual RVs points. For a given (Mass; Period), the distribution of all the standard deviations (corresponding to differents $T_0$) obtained is gaussian. We compute then the average value of the distribution of the virtual standard deviations. We consider that a planet with a given (Mass; Period) is detectable if the standard deviation of the real RV values is smaller than the average value of the virtual standard deviations. We determine the confidence level (or detection probability) associated to such an orbit by comparing the standard deviation of the virtual distribution with the difference between the standard deviation of the real RV measurements and the average value of the virtual standard deviations.
 
In practice, for a given object, we explore 200 periods in the range 0.5 to 1000 days, and 100 planet masses in the range (Mmin; 100\,\Mjup) where Mmin corresponds to the achievable mass given the measured uncertainty. For a given (Mass; Period), we explore 1000 $T_0$. We checked that increasing the number of periods, or planet masses and or initial $T_0$ does not significantly impact the results.

For each (Mass; Period) couple, we obtain thus a detection probability. In a (Mass; Period) diagram, we can then identify the domain where a planet with a given mass and period should be detectable if present, with a given level of confidence. This defines then a domain in which we can exclude the presence of a planet with a given level of confidence. We will consider two levels of confidence: 1$\sigma$, (\ie a 68.2\,$\%$ probability) and 3$\sigma$ (\ie 99.7\,$\%$ probability).  


 %
\subsection{Sensitivity of the survey and first constrains on early type stars}
The sensitivity of our survey is a consequence of the number of data available and on the temporal sampling of the data. We kept only those stars (107 objects), found to be either constant or variable, for which we got more than 12 data points (6 epochs). Also, given the data at hand, we limited the range of periods investigated between 1\,days and a few hundred days. We report in
Table 3 the achieved limits (68.2\,$\%$ and 99.7\,$\%$ probabilities) obtained for each of the 107 stars considering three periods: 3, 10 and 100 days.

\begin{figure*}
  \centering
  \includegraphics[width=0.9\hsize]{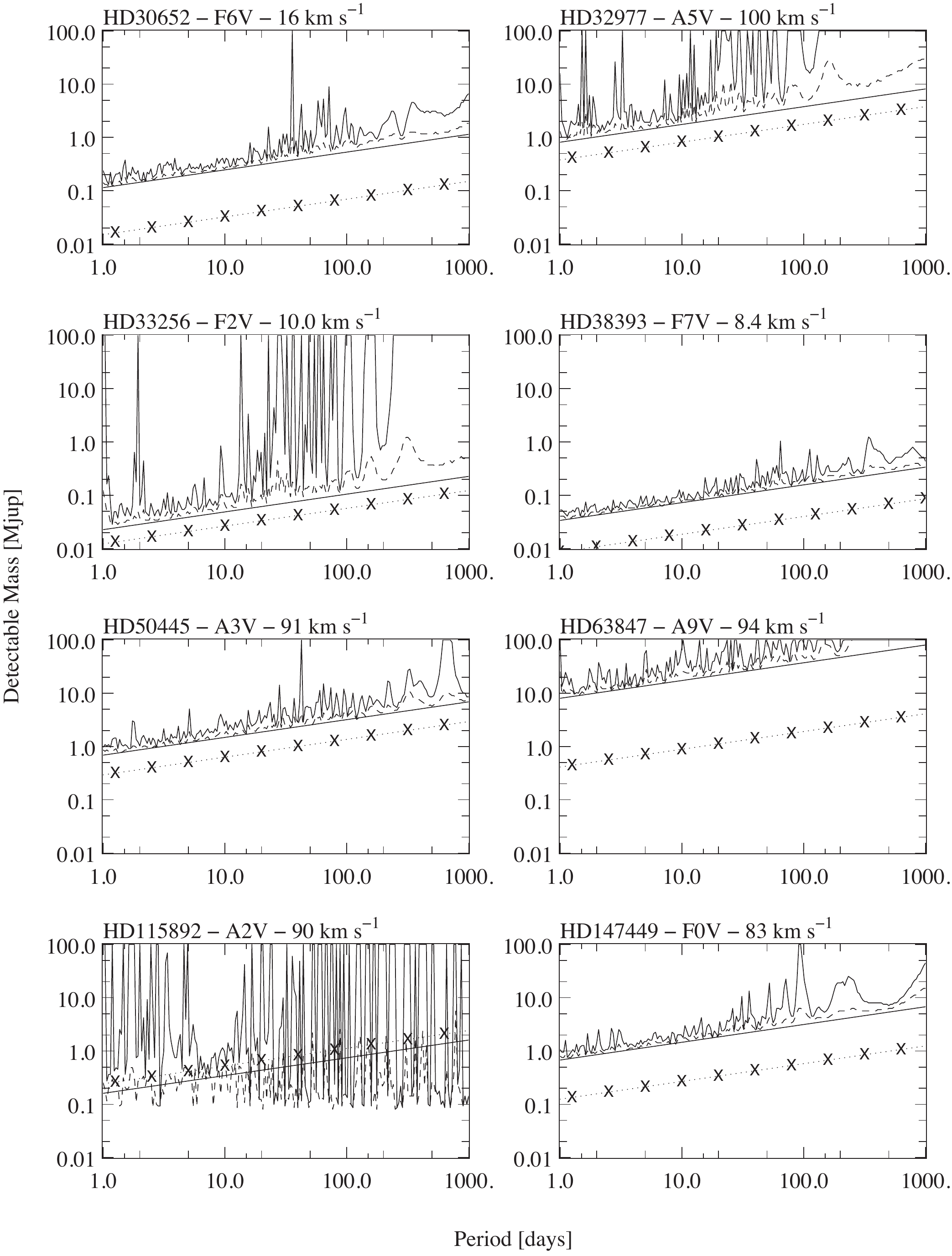}
  \caption{Detection limits. X-axis: periods (days). Y-axis: Detection limit (M/\Mjup). Curve: detection limits actually achieved in the present survey; plain curves correspond to 99.7 $\%$ detection probability, and broken curves to 68.2 $\%$ probability. Note that the RV data were beforehand averaged over one day.
Straight line: achievable detection limits assuming a 3$\times$ rms threshold for the planet amplitude. Line with crosses: achievable detection limits assuming a  3$\times$ uncertainty threshold for the planet amplitude.}
  \label{limdet}
\end{figure*}

We give also in Fig.\,\ref{limdet} examples of achieved detection limits (68;2\,$\%$ and 99.7\,$\%$ probabilities) as estimated with the previously described simulations. We have also plotted on the figures the achievable detection limits taking into acount the jitter, as defined in the previous section, as well as  the detection limits corresponding to the measured uncertainties. We recall that the last two cases (achievable limits) do not take into account the actual temporal sampling of the data, conversely to the detection limits computed with our virtual realisations. 

When enough data are available, the actual detection limits fall close to the achievable limits obtained assuming the $3 \times {\rm RV}$ rms threshold for the amplitude of RV variations, as can be seen in Fig.\,\ref{limdet}. This justifies the choice of the threshold adopted in the previous section to estimate the achievable limits.


Note that we sometimes endup with high detection limits that fall outside the investigated range of masses, \ie $\geq$ 100\,\Mjup~when we consider a period of 100 days whereas the detection limits for a 3 or 10 days period are close to the achievable limits. This corresponds to cases where the temporal sampling is not adapted to explore such a long period (see for instance the case of HD\,33256, in Fig.\,\ref{limdet}). It sometimes, but much less frequently also happens that the detection limit for a 10 days fall outside the investigated range of masses whereas the detection limits for a 3 days period is close to the achievable limit. In fact, our survey allows to search mostly for short-period planets (typ. a few days); 10--100 day periods are not always sampled enough to get interesting results (especially on early-type stars) and $\geq$ 100-day periods are not properly sampled to get interesting results. We will therefore in the following discuss only periods $\leq$ 100 days. Finally, one has to note that in some cases, we get a 99.7 $\%$ probability detection limit out of the investigated range, where as the 68.2 $\%$ limit fall well into the investigated range. This happens generally when the number of data is the lowest: 12 or 14.

\addtocounter{table}{1}

To study the impact of \bv~on the present results, we computed the percentage of stars for which the achieved detection limits (68.2\,$\%$ and 99.7\,$\%$ probabilities) fall in the planetary or BD domains per bin of \bv, considering a 3-day, a 10-day and a 100-day periods. We also computed the median of the achieved detection limits (considering 68.2\,$\%$ and 99.7\,$\%$ probabilities) per bin of \bv~for such periods. The results are given in Table\,\ref{achievedlimdet}. In order to allow comparison between the limits obtained with these two probabilities and with the achievable ones, we considered for the computation of the median values only those stars for which both the 68.2\,$\%$ and 99.7\,$\%$ probabilities detection limits fall into the investigated range of masses. Finally, one has to note that for the earliest type stars, the number of objects per bin is quite small, so one has to be very cautious with the associated statistics. We can see that:
\begin{itemize}
  \item[-] if we consider a 3-day period, the achieved limit at 1 (resp. 3) $\sigma$ fall in the planetary domain for 90\,$\%$ (resp. 81 \,$\%$) of the stars. This percentage is quite comparable to the one found in Sect.~3. It increases from 75 (resp. 25)\,$\%$ for the earliest type stars to 100 (resp. 100) \,$\%$ for stars with \bv~larger than 0.3. Also, the median of the achieved limits at 1 (resp. 3) $\sigma$ decreases from 7 (resp. 7) \Mjup~for the earliest-type stars to 0.08 (resp. 0.3) \Mjup~for the latest-type stars. Moreover, the steep step seen in Sect.\,\ref{results} in the detectable masses at \bv~= 0.3 is also clear.
 \item[-] if we consider a 10-day period, the achieved limit at 1 $\sigma$ (resp. 3) fall in the planetary domain for 82\,$\%$ (resp. 67\,$\%$) of the stars. This percentage increases from 50 (resp. 25)\,$\%$ for the earliest type stars to 100 (resp. 100) \,$\%$ for stars with \bv~larger than 0.4, with however an exception on the [0.2; 0.3] range where it decreases back to 50\,$\%$. Also, the median of the achieved limits at 1 (resp. 3) $\sigma$ decreases from 12 (resp. 24) \Mjup~for the earliest-type stars to 0.1 (resp. 0.7) for the latest-type stars, with however an exception in the range [0.2; 0.3] range as regards the 99.7\,$\%$ probability. Again the steep step is observed at \bv~= 0.3.
\item[-] if we finally consider a 100-day period, the achieved limit at 1 (resp. 3) $\sigma$ fall in the planetary domain for 54 (resp. 35) $\%$ of the stars. We note that this percentage is smaller than the one obtained in Sect.~3 and attribute the discrepancy to the actual temporal sampling and the relatively small amount of targets investigated yet. The median of the achieved limits at 1 (resp. 3) $\sigma$ decreases from 19 (resp. 34) \Mjup~for stars with \bv~$\geq$ 0.0 to 1. (resp. 1.9) \Mjup~for the latest-type stars. Again the steep step is observed at \bv~= 0.3.
\end{itemize}

\begin{table*}[htp]
 \caption{Percentage of stars for which the achieved detection P=68.2$\%$ or P=99.7$\%$ limits fall in the planetary or BD/planet domains. Three periods are considered: 3, 10 and 100 days. Median values of the {\it achieved} detection limits, expressed in Jupiter mass, per bin of \bv. For the these median values, we took into acount only the stars for which both the P=68.2$\%$ and P=99.7$\%$ limits fall into the investigated companion mass range. The corresponding number of stars is given as well as the median achievable limits for those stars.}
  \label{achievedlimdet}
  \begin{center}
    \begin{tabular}{l l l l l l l l}
 \hline
\bv		                     & [$-$0.1; 0]& [0; 0.1] & [0.1; 0.2]  & [0.2; 0.3] & [0.3; 0.4] & [0.4; 0.5] & [0.5; 0.6]\\
\hline
\hline
Number of stars   (whole sample)             &  4   &19  &21       &10      &12         &24     &17  \\
\hline
$P=3$\,days                                  &     &      &         &         &         &       &  \\
Percentage of limits in the planet domain & 75;25      & 74;47      & 95;61        & 100;60       &100;100        &100;100      & 100;100\\
   (P=68.2$\%$;P=99.7$\%$)                           &     &      &         &         &         &       &  \\
Percentage of limits in the BD/planet domain          & 100;100      & 100;100      & 100;100       & 100;100       &100;100        &100;100      & 100;100\\
  (P=68.2$\%$;P=99.7$\%$)                           &     &      &         &         &         &       &  \\
\hline
$P=10$\,days  &     &      &         &          &         &       &  \\
Percentage of limits in the planet domain    & 50;25  & 58;42      & 80;52        & 50;30       &92;66         &100;100        &100;100\\
  (P=68.2$\%$;P=99.7$\%$)                           &     &      &         &         &         &       &  \\
Percentage of limits in the BD/planet domain        & 100;100  & 100;52      &100;95         & 100;64        &100;92         &100;92        &100;100\\  (P=68.2$\%$;P=99.7$\%$)                           &     &      &         &         &         &       &  \\
\hline
$P=100$\,days  &     &      &         &         &     &       &  \\
Percentage of limits in the planet domain       & 25;0      & 16;10      & 28;10         & 30;10        &58;33         &92;55        &94;47 \\
  (P=68.2$\%$;P=99.7$\%$)                           &     &      &         &         &         &       &  \\
 Percentage of limits in the BD/planet domain       & 75;25   & 74;83      & 95;48       & 100;40     &100;58       &100;59        &100;100 \\
  (P=68.2$\%$;P=99.7$\%$)                           &     &      &         &         &         &       &  \\
\hline
\hline
$P=3$\,days                                  &     &      &         &         &         &       &  \\
Number of stars                              & 4   & 17   & 17      & 8       &11       &22     &17       \\
Achieved detection limit (P=68.2$\%$)          & 6.9 & 5.3  & 5.1     & 3.2     &0.5      &0.25   &0.08  \\
Achieved detection limit (P=99.7$\%$)          & 6.9 & 12.1 & 7.3     & 6.3     &1.0      &0.3    &0.3  \\
Achievable detection limit                   & 6.9 & 5.5  & 4.2     & 2.5     &0.4      &0.2     &0.06  \\
\hline
$P=10$\,days  &     &      &         &          &         &       &  \\
Number of stars                              &  4   &14  &20       &7      &11         &22     &17  \\
Achieved detection limit (P=68.2$\%$)          & 11.7 &8.0 &7.2      &4.4    &0.9        &0.4    &0.1  \\
Achieved detection limit (P=99.7$\%$)          & 24.0 &11.8&10.2      &17.4   &2.0        &0.8    &0.7 \\
Achievable detection limit                   & 10.3 &6.9 &5.6      &4.1    &0.6        &0.3    &0.1  \\
\hline
$P=100$\,days  &     &      &         &         &     &       &  \\
Number of stars                   & 1           &6    &11   &5    &8   &13 &12  \\
Achieved detection limit (P=68.2$\%$) &   (10.0)  &18.7 &17.3 &14.2 &5.2 &1.1&1.0  \\
Achieved detection limit (P=99.7$\%$) &   (17.6)  &34.4 &32.9 &33.8 &20.4&2.4&1.9  \\
Achievable detection limit          &    (8.5)  &15.0 &10.5 &8.9  &3.0 &0.8&0.6  \\

     \hline
  \end{tabular}
  \end{center}
\end{table*}

Finally, we give the probability {\it not} to detect planets of a given mass and with a given period (3, 10, 200, 500 days) around stars with a given spectral type and \vsini. The results are summarized in Table\,\ref{stats}. We see that as expected, for a given probability, the limits globally decrease with increasing B$-$V and decreasing \vsini. 

\begin{table}[t]
  \caption{Detection limit for 50\,$\%$  and 90\,$\%$ for different periods. The last column give the number of stars considered to estimate these detection limits. Note that only stars with more than 12 measurements (6 epochs) were considered. Binaries were excluded. Also note that numbers outside the planetary mass domain ($>13$\,\Mjup) are not given.}
  \label{stats}
  \begin{center}
    \begin{tabular}{l l l l l }
      \hline
      \hline
      ST, \vsini                       & Period & 50\,$\%$ & 90\,$\%$  & N-st. \\
                                                 & [days] & [\Mjup]  & [\Mjup]   & \\
      \hline
      early A, \vsini~$\leq$ 70\,\kms	                & 3      & 1.2      & --       & 2 \\
                                      			& 10     & 1.7      & --      & 2 \\
                                      			& 200    & 4.7      & --      & 2 \\
                                      			& 500    & 6.4      & --      & 2 \\
      early A, \vsini~70-130\,\kms		& 3      & 4.6      & 7.2       & 9 \\
                                      			& 10     & 6.9      & 10.7      & 9 \\
      \hline
      A, \vsini~$\leq$ 70\,\kms      		& 3      & 3.5      & --       & 2 \\
      A, \vsini~70-130\,\kms          		& 3      & 4.5      & 6.3       & 10 \\
                                      			& 10     & 6.7      & 9.5       & 10 \\
      A, \vsini~$\geq$ 130\,\kms      	& 3      & 11.2      &           & 13 \\
      \hline
      F, \vsini~$\leq$ 15\,\kms       	& 3      & 0.1      & 0.8       & 28 \\
                                      			& 10     & 0.2      & 1.2       & 28 \\
                                      			& 200    & 0.4      & 3.2       & 25 \\
                                      			& 500    & 0.6      & 2.7       & 22 \\
      F, \vsini~15-60\,\kms           		& 3      & 0.6      & 0.8       & 13 \\
                                      			& 10     & 0.9      & 1.2       & 13 \\
                                      			& 200    & 2.8      & 3.3       & 12 \\
                                      			& 500    & 3.8      & 4.5       & 12 \\
      F, \vsini~$\geq$ 60\,\kms       	& 3      & 1.9      & 10.2       & 10 \\
                                      			& 10     & 2.8      &           & 10 \\
                                      			& 200    & 7.5      &           & 9 \\
                                      			& 500    & 10.1     &           & 9 \\
      \hline
    \end{tabular}
  \end{center}
\end{table}


Obviously, the statistics provided by our survey is still poor on early type stars, and still very limited on the latest type stars. Concerning the latter, we note that if we consider the 41 objects with \bv~$\geq$ 0.4, \ie well beyond the instability strip, we find that less than 24\,$\%$ of stars host planets with masses equal to 0.5\,\Mjup~or more; less than 5\,$\%$ host planets with masses equal to 1\,\Mjup~or more on a 3-day period. For a 10-day period, we find that less than 42\,$\%$ host planets with masses $\geq$ 0.5\,\Mjup, and less than 20\,$\%$) host planets with masses $\geq$ 1\,\Mjup. The comparison between achieved and achievable detection limits shows that there is still room to significantly improve those statistics (thanks to new data points). 

The present statistics certainly does not allow quantitative comparisons with late type dwarfs, which have been surveyed by several groups since more than 10 years, or with giant or sub giant stars, because in that case of the lack of data for both massive dwarfs and (sub-)giants. 

Concerning the presence or absence of hot Jupiters around massive stars, we note that the planets found so far in our survey are located at about 0.7 AU or more from a 1.4\,\Msun~star; this separation corresponds to that of the closest planet found around giant stars. We also detected recently a planet orbiting at 0.6\,AU from a dwarf with a similar mass in the northern hemisphere (\cite{desort08b}). Due to the still limited amount of data available, it should not be concluded however that there are no planets closer to massive dwarfs. We also recall that a few short period planets have been found around 1.4\,\Msun~stars through transits. 

\section{Conclusion}
Based on the observation of a large number of A--F type stars (170), we have been able to measure their jitters, and derive for the first time estimations of the detection limits that can be expected on average on those stars with \bv~in that range [$-$0.1; 0.6] (once previously known $\delta$\,Scuti and $\gamma$\,Doradus stars are removed) for 3 periods: 3, 10 and 100-day periods. We have shown that at the precision provided by {\small HARPS}, most of the stars are variable in RV, and the impact of the RV jitter, due to either spots or pulsations is generally non negligeable on planet detectability. However, assuming that planets are detectable if the amplitude of the induced RV variation is larger than 3$\times {\rm rms}$ (this threshold defines the achievable detection limits, which depends on the star and the spectrograph used), we have shown that even when taking into account the jitter, giant planets can still be found around these stars in most cases. This is not only true for the stars with \bv~$\geq$ 0.3 for which we can find either short or long period planets, with masses as small as 0.02\,\Mjup~(case of short period) for the latest type stars but also for dwarfs with \bv~$\leq$ 0.3: for such stars short-period planets can still be found around those with relatively low projected rotational-velocity and low level of activity, with masses down to 0.5\,\Mjup~(best case). This survey has allowed to identify for the first time the stars that are best suited for further searches for planets around massive dwarfs.

We have also shown that given the data available, the present survey is sensitive to short-period planets (hot Jupiters) and only partially sensitive to larger periods (up to 100\,days). We found in particular one 2-planet system with periods larger than 100\,days around one late-type star. Whenever possible (107 stars), we computed for each star the detection limits actually achieved and showed that when enough data are available, the achieved detection limit is set by the $3 \times {\rm rms}$ threshold. We indeed reached such limits for early-type as well as for late-type stars. We finally derive first estimates of the presence of short-period planets around these A--F stars. We showed for instance that less than 5\,$\%$ of the latest-type stars (\bv~$\geq$ 0.4) host P=3 days-period planets with masses 1\,\Mjup~or more. Such statistics are not constraining enough to allow interesting comparisons with later type stars or with model predictions; but as soon as more data become available, the statistics can be straightforwardy improved.

Finally, we note that to compute these detection limits, we did not try to average out the spectra over timescales associated to the frequencies of intrinsic stellar variations. This approach would of course allow to decrease significantly the detectable masses. As it would require lots of telescope time, it should be probably kept for stars with the highest scientific interest.


\begin{acknowledgements}
  We acknowledge financial support from the French Programme National de Plan\'etologie ({\small PNP, INSU}). We also acknowledge
 support from the French National Research Agency (ANR) through project
 grant NT05-4$\_$44463.

  These results have made use of the SIMBAD database, operated at CDS, Strasbourg, France.

  We also thank G\'erard Zins and Sylvain C\`etre for their help in implementing the SAFIR interface, Sylvain C\`etre also for performing 
some of the observations, and P. Rubini for his help on the lay out of the paper.
\end{acknowledgements}

\longtab{1}{
  \small{
    \begin{center}
      \begin{longtable}{l l l l l l l l l l l l l l}
	\caption{Stars properties and measurements. ``RV rms'' (resp. ``span rms'') stands for the rms of the measured radial velocities (resp. bisector velocity-spans); ``RV amp'' (resp. ``Span amp.'') stands for amplitude of the measured radial velocities (resp. bisector velocity-spans); ``RV unc.'' (resp. ``Span unc.'') stands for the average uncertainties associated to the RV (resp. bisector velocity-spans) data. Bisector flags: G: good quality; B: bad quality; A: acceptable quality. Binary types: X refers to stars identified as binaries based on a high $\chi^2$ ($\geq$ 10); B refers to binaries identified via a flat or a composite bisector, and V refers to stars regarded as binary candidates, based on the sole amplitude of their RV variations (see text).}
	\label{sample}
	\\
	\hline
	HD & ST & \bv & \vsini & Time   & RV    & RV    & RV    & Span  & Span  & Span  & Bis. & Variabl. & Bin. \\
	   &    &     &        & bs  l  & rms   & unc   & ampl  & rms   & unc   & ampl  & Flag &          & \\
	   &    &     & (\kms) & (days) & (\ms) & (\ms) & (\ms) & (\ms) & (\ms) & (\ms) &      &          & \\
	\hline
	\object{HD\,693}   & F5V & 0.487 & 10  & 389.1  & 2    & 1    & 6    & 2    & 1    & 7     & G & V & \\
	\object{HD\,2696}  & A3V & 0.128 & 150 & 389.1  & 279  & 166  & 847  &      &      &       &   & C & \\
	\object{HD\,2834}  & A0V & 0.018 & 130 & 279.2  & 9408 & 164  & 29717&      &      &       &   & V & V\\
	\object{HD\,2884}  & B9V & -0.06 & 170 & 603.2  & 587  & 496  & 2141 &      &      &       &   & C & \\
	\object{HD\,2885}  & A2V & 0.147 & 40  & 820.7  & 9331 & 19   & 34510&      &      &       & B & V & X\\
	\object{HD\,3003}  & A0V & 0.038 & 115 & 603.2  & 135  & 83   & 491  &      &      &       &   & C & \\
	\object{HD\,4247}  & F0V & 0.35  & 35  & 842.8  & 26   & 11   & 94   & 62   & 49   & 216   & G & V & \\
	\object{HD\,4293}  & A7V & 0.297 & 125 & 837.8  & 91   & 99   & 270  & 4109 & 1281 & 16151 & B & C & \\
	\object{HD\,7439}  & F5V & 0.448 & 8   & 453.8  & 10   & 1    & 22   & 24   & 2    & 63    & G & V & \\
	\object{HD\,9672}  & A1V & 0.066 & 195 & 453.8  & 169  & 300  & 527  &      &      &       &   & C & \\
	\object{HD\,11262} & F6V & 0.523 & 5   & 840.8  & 2500 & 1    & 5014 & 30   & 2    & 64    & G & V & B\\
	\object{HD\,12311} & F0V & 0.29  & 22  & 387.9  & 3674 & 80   & 10137& 3113 & 757  & 12217 & G & V & B\\
	\object{HD\,13555} & F5V & 0.457 & 9   & 821.8  & 13   & 1    & 42   & 20   & 3    & 54    & G & V & \\
	\object{HD\,14943} & A5V & 0.213 & 111.5& 835.8 & 141  & 50   & 667  & 745  & 484  & 6364  & A? & V & \\
	\object{HD\,15008} & A3V & 0.034 & 180 & 1165.9 & 367  & 274  & 1210 &  &  &  &  & C & \\
	\object{HD\,17848} & A2V & 0.101 & 220 & 842.8  & 294  & 194  & 1046 &  &  &  &  & C & \\
	\object{HD\,18978} & A4V & 0.163 & 120 & 1161.8 & 185  & 88   & 880  & 13988& 1317 & 105268& B & V & \\
	\object{HD\,19107} & A8V & 0.193 & 170 & 389    & 154  & 176  & 434  &  &  &  &  & C & \\
	\object{HD\,19545} & A3V & 0.166 & 80  & 821.9  & 1490 & 33   & 3962 & 623  & 244  & 2389  & A & V & B\\
	\object{HD\,21882} & A5V & 0.205 & 255 & 382.9  & 400  & 350  & 1332 &  &  &  &  & C & \\
	\object{HD\,25457} & F5V & 0.516 & 25  & 368.1  & 28   & 3    & 93   & 28   & 8    & 82    & G & V & \\
	\object{HD\,25490} & A1V & 0.032 & 65  & 5      & 74   & 71   & 214  &  &  &  &  & C & \\
	\object{HD\,29488} & A5V & 0.147 & 115 & 665.2  & 183  & 72   & 712  & 3160 & 855  & 23124 & B & V & \\
	\object{HD\,29875} & F2V & 0.342 & 45  & 325    & 434  & 7    & 1100 & 1108 & 36   & 2551  & G & V & \\
	\object{HD\,29992} & F3V & 0.391 & 100 & 638.2  & 335  & 30   & 855  & 319  & 226  & 1386  & G & V & \\
	\object{HD\,30652} & F6V & 0.484 & 16  & 500.7  & 13   & 2    & 54   & 19   & 7    & 90    & G & V & \\
	\object{HD\,30739} & A1V & 0.01  & 195 & 328    & 941  & 631  & 3380 &  &  &  &  & C & \\
	\object{HD\,31746} & F3V & 0.442 & 11.4& 295    & 18   & 2    & 60   & 45   & 6    & 144   & G & V & \\
	\object{HD\,32743} & F2V & 0.421 & 50  & 663.2  & 13   & 5    & 44   & 145  & 42   & 441   & A & V & \\
	\object{HD\,32977} & A5V & 0.118 & 100 & 175.7  & 70   & 47   & 279  & 272  & 340  & 990   & A/B & C & \\
	\object{HD\,33256} & F2V & 0.455 & 10  & 326    & 4    & 2    & 14   & 5    & 4    & 16    & G & V & \\
	\object{HD\,33262} & F7V & 0.526 & 30  & 665.1  & 17   & 3    & 45   & 60   & 7    & 173   & G & V & \\
	\object{HD\,37306} & A2V & 0.051 & 130 & 670.2  & 275  & 180  & 1127 &  &  &  &  & C & \\
	\object{HD\,38393} & F7V & 0.481 & 8.4 & 1165.9 & 4    & 1    & 19   & 5    & 2    & 22    & G & V & \\
	\object{HD\,38678} & A2V & 0.104 & 200 & 338.1  & 723  & 429  & 1942 &  &  &  &  & C & \\
	\object{HD\,39014} & A7V & 0.217 & 205 & 665.2  & 512  & 173  & 2376 &  &  &  &  & V & \\
	\object{HD\,39060} & A3V & 0.171 & 130 & 670.1  & 287  & 39   & 996  & 883 & 423 & 3702 & A & V & \\
	\object{HD\,40136} & F1V & 0.337 & 18  & 663.2  & 10   & 3    & 36   & 19 & 7 & 77 & G & V & \\
	\object{HD\,41695} & A0V & 0.046 & 250 & 280.2  & 595  & 628  & 2046 &  &  &  &  & C & \\
	\object{HD\,41742} & F4V & 0.493 & 26.3& 450.9  & 673  & 5    & 2030 & 44 & 17 & 148 & G & V & B\\
	\object{HD\,43940} & A2V & 0.139 & 247.5& 282.2 & 597  & 513  & 1939 &  &  &  &  & C & \\
	\object{HD\,46089} & A3V & 0.185 & 110 & 0.1    & 1060 & 64   & 2618 & 3148 & 687 & 8811 & A & V & \\
	\object{HD\,49095} & F7V & 0.491 & 7   & 337.9  & 3    & 1    & 13   & 4 & 2 & 21 & G & V & \\
	\object{HD\,49933} & F2V & 0.396 & 12  & 29     & 29   & 2    & 85   & 83 & 4 & 237 & G & V & \\
	\object{HD\,50445} & A3V & 0.183 & 90.7& 670.1  & 66   & 36   & 248  & 301 & 295 & 1672 & A & C & \\
	\object{HD\,54834} & A9V & 0.312 & 26.9& 0.9    & 1183 & 11   & 2839 & 127 & 657 & 312 & G & V & B\\
	\object{HD\,56537} & A3V & 0.106 & 140 & 666.2  & 180  & 75   & 639  & 3198 & 1090 & 15172 & A & V & \\
	\object{HD\,59984} & F5V & 0.54  & 15  & 29.9   & 3    & 1    & 10   & 30 & 24 & 107 & B & V & \\
	\object{HD\,60532} & F6V & 0.521 & 10  & 667    & 26   & 1    & 109  & 4 & 2 & 22 & G & V & \\
	\object{HD\,60584} & F6V & 0.468 & 38.9& 663.1  & 23   & 9    & 83   & 62 & 44 & 215 & G & V & \\
	\object{HD\,63847} & A9V & 0.31  & 94  & 339    & 794  & 65   & 3145 & 997 & 565 & 3835 & A & V & \\
	\object{HD\,66664} & A1V & 0.018 & 175 & 32.8   & 542  & 450  & 1695 & 0 & 0 & 0 &  & C & \\
	\object{HD\,68146} & F7V & 0.488 & 8   & 666.1  & 4    & 1    & 16   & 5 & 3 & 27 & G & V & \\
	\object{HD\,68456} & F5V & 0.437 & 12  & 212.3  & 1236 & 2    & 3613 & 70 & 4 & 177 & G & V & B\\
	\object{HD\,71155} & A0V & -0.012 & 115& 337.9  & 480  & 227  & 2288 &  &  &  &  & V & \\
	\object{HD\,73262} & A1V & 0.003 & 265 & 30.9   & 1101 & 777  & 3843 &  &  &  &  & C & \\
	\object{HD\,74591} & A6V & 0.2   & 115 & 338    & 171  & 80   & 639  & 726 & 901 & 3137 & B & V & \\
	\object{HD\,74873} & A1V & 0.12  & 10  & 28.9   & 1534 & 114  & 5167 & 0 & 0 & 0 &  & V & \\
	\object{HD\,75171} & A9V & 0.217 & 93.3& 388    & 294  & 57   & 934  & 713 & 443 & 2021 & A & V & \\
	\object{HD\,76653} & F6V & 0.481 & 11  & 282.1  & 13   & 1    & 45   & 14 & 3 & 55 & G & V & \\
	\object{HD\,77370} & F3V & 0.417 & 95  & 667.1  & 37   & 15   & 134  & 136 & 71 & 504 & G & V & \\
	\object{HD\,82165} & A6V & 0.216 & 232.8& 32    & 541  & 276  & 2046 &  &  &  &  & C & \\
	\object{HD\,83446} & A5V & 0.173 & 155 & 670.1  & 274  & 83   & 1152 &  &  &  &  & V & \\
	\object{HD\,88955} & A2V & 0.051 & 105 & 99.8   & 115  & 69   & 450  &  &  &  &  & C & \\
	\object{HD\,89328} & A8V & 0.329 & 302.8& 568.3 & 340  & 93   & 956  &  &  &  &  & V & \\
	\object{HD\,90132} & A8V & 0.251 & 270 & 28.9   & 433  & 221  & 1647 &  &  &  &  & C & \\
	\object{HD\,91324} & F6V & 0.5   & 8   & 665.1  & 3    & 1    & 13   & 6 & 3 & 25 & G & V & \\
	\object{HD\,91889} & F7V & 0.528 & 6   & 665    & 4    & 1    & 17   & 2 & 1 & 9 & G & V & \\
	\object{HD\,93372} & F6V & 0.51  & 11.3& 638.1  & 4    & 2    & 14   & 8 & 5 & 28 & G & C & \\
	\object{HD\,94388} & F6V & 0.48  & 8   & 31.9   & 15   & 1    & 51   & 27 & 3 & 79 & G & V & \\
	\object{HD\,96819} & A1V & 0.069 & 230 & 5      & 463  & 442  & 1520 &  &  &  &  & C & \\
	\object{HD\,97244} & A5V & 0.209 & 75  & 106.8  & 84   & 45   & 298  & 301 & 360 & 1246 & A & C & \\
	\object{HD\,97603} & A4V & 0.128 & 165 & 32.9   & 315  & 170  & 987  & 0 & 0 & 0 &  & C & \\
	\object{HD\,99211} & A9V & 0.216 & 130 & 29.9   & 135  & 59   & 474  & 433 & 592 & 1796 & A & V & \\
	\object{HD\,99453} & F7V & 0.495 & 5   & 32     & 253  & 1    & 784  &  &  &  &  & V & X\\
	\object{HD\,100563} & F5V & 0.48  & 14  & 564.3 & 3    & 2    & 10   & 12 & 4 & 35 & G & C & \\
	\object{HD\,101198} & F7V & 0.52  & 5   & 662.1 & 40   & 1    & 102  & 2 & 1 & 5 & G & V & \\
	\object{HD\,102124} & A4V & 0.174 & 130 & 29.9  & 297  & 114  & 1195 & 2230 & 1928 & 7132 & B & V & \\
	\object{HD\,102647} & A3V & 0.09  & 115 & 106.8 & 111  & 44   & 426  & 419 & 454 & 2305 & A & V & \\
	\object{HD\,104731} & F6V & 0.417 & 20  & 29.9  & 23   & 2    & 75   & 129 & 6 & 445 & G & V & \\
	\object{HD\,105850} & A1V & 0.055 & 122 & 569.3 & 181  & 107  & 606  &  &  &  &  & C & \\
	\object{HD\,106661} & A3V & 0.068 & 175 & 32    & 358  & 373  & 1272 &  &  &  &  & C & \\
	\object{HD\,109085} & F2V & 0.388 & 81  & 638   & 22   & 15   & 77   & 86 & 82 & 329 & G & C & \\
	\object{HD\,109787} & A2V & 0.049 & 330 & 32.9  & 829  & 509  & 3038 &  &  &  &  & C & \\
	\object{HD\,110411} & A0V & 0.076 & 140 & 32.9  & 851  & 415  & 2748 &  &  &  &  & V & \\
	\object{HD\,111998} & F5V & 0.493 & 28.5& 638.1 & 40   & 5    & 144  & 35 & 18 & 128 & G & V & \\
	\object{HD\,112934} & A9V & 0.298 & 70  & 73.8  & 857  & 43   & 2877 & 580 & 289 & 1950 & G & V & B\\
	\object{HD\,114642} & F6V & 0.46  & 13  & 105.7 & 49   & 2    & 194  & 118 & 4 & 448 & G & V & \\
	\object{HD\,115892} & A2V & 0.068 & 90  & 101.8 & 59   & 29   & 232  &  &  &  &  & V & \\
	\object{HD\,116160} & A2V & 0.045 & 205 & 97.7  & 4080 & 437  & 10844&  &  &  &  & V & V\\
	\object{HD\,116568} & F3V & 0.415 & 40  & 74.9  & 1243 & 8    & 2751 & 154 & 36 & 592 & G & V & B\\
	\object{HD\,118098} & A3V & 0.114 & 205 & 105.9 & 395  & 263  & 1533 &  &  &  &  & C & \\
	\object{HD\,124850} & F7V & 0.511 & 15  & 5     & 39   & 2    & 129  & 88 & 6 & 278 & G & V & \\
	\object{HD\,125276} & F7V & 0.518 & 5   & 182.8 & 1    & 1    & 4    & 2 & 1 & 8 & G & C\\
	\object{HD\,126248} & A5V & 0.124 & 185 & 280.2 & 602  & 204  & 2482 &  &  &  &  & V\\
	\object{HD\,128020} & F7V & 0.506 & 5   & 73.8  & 2    & 1    & 6    & 3 & 2 & 10 & G & C\\
	\object{HD\,128167} & F3V & 0.364 & 8   & 6     & 21   & 3    & 70   & 56 & 5 & 169 & G & V\\
	\object{HD\,128898} & F1V & 0.256 & 15  & 0.1   & 64   & 2    & 189  & 41 & 6 & 121 & G & V\\
	\object{HD\,129422} & A7V & 0.308 & 200 & 113.9 & 562  & 196  & 1742 &  &  &  &  & V\\
	\object{HD\,129926} & F0V & 0.315 & 110 & 73.9  & 347  & 26   & 1074 & 642 & 214 & 2358 & G & V\\
	\object{HD\,130109} & A0V & -0.005& 265 & 31    & 1319 & 945  & 4253 &  &  &  &  & C\\
	\object{HD\,132052} & F0V & 0.318 & 105 & 182.6 & 139  & 31   & 496  & 495 & 253 & 1490 & G & V\\
	\object{HD\,133469} & F6V & 0.489 & 24.3& 182.6 & 17   & 5    & 60   & 32 & 17 & 104 & G & V\\
	\object{HD\,135379} & A3V & 0.088 & 60  & 841.8 & 31   & 22   & 105  & 109 & 153 & 435 & G & C\\
	\object{HD\,135559} & A4V & 0.181 & 125 & 389   & 505  & 87   & 2063 & 1626 & 1166 & 6894 & B & V\\
	\object{HD\,138763} & F7V & 0.577 & 7   & 623.2 & 56   & 2    & 200  & 54 & 4 & 217 & G & V\\
	\object{HD\,139211} & F6V & 0.505 & 7   & 182.7 & 3    & 1    & 12   & 3 & 2 & 12 & G & V\\
	\object{HD\,141513} & A0V & -0.036& 85  & 389   & 147  & 55   & 599  &  &  &  &  & V\\
	\object{HD\,141851} & A3V & 0.135 & 185 & 114.9 & 604  & 482  & 1899 &  &  &  &  & C\\
	\object{HD\,142139} & A3V & 0.087 & 110 & 5     & 45   & 45   & 142  & 234 & 327 & 681 & A & C & \\
	\object{HD\,142629} & A3V & 0.129 & 85  & 386.9 & 2207 & 25   & 6526 &  &  &  & G & V & X\\
	\object{HD\,145689} & A4V & 0.159 & 100 & 112.9 & 198  & 62   & 604  & 807 & 618 & 2460 & B & V & \\
	\object{HD\,146514} & A9V & 0.326 & 145 & 389   & 820  & 134  & 2368 & 2559 & 1861 & 7544 & B & V & \\
	\object{HD\,146624} & A0V & 0.008 & 30  & 724.9 & 10   & 15   & 33   & 51 & 57 & 167 & G & C & \\
	\object{HD\,147449} & F0V & 0.338 & 83  & 389   & 65   & 19   & 265  & 216 & 127 & 996 & G & V & \\
	\object{HD\,153363} & F3V & 0.407 & 27  & 279.2 & 204  & 10   & 590  & 236 & 41 & 637 & G & V & \\
	\object{HD\,156751} & A5V & 0.248 & 93.8& 5.9   & 474  & 80   & 1359 & 1281 & 607 & 3866 & B & V & \\
	\object{HD\,158094} & B8V & -0.104& 255 & 386   & 1876 & 539  & 5639 &  &  &  &  & V & V\\
	\object{HD\,158352} & A8V & 0.237 & 165 & 388   & 1399 & 90   & 3968 &  &  &  &  & V & V\\
	\object{HD\,159170} & A5V & 0.187 & 225 & 114.8 & 636  & 344  & 2541 &  &  &  &  & C & \\
	\object{HD\,159492} & A7V & 0.195 & 60  & 446.9 & 106  & 20   & 371  & 429 & 108 & 1333 & G & V & \\
	\object{HD\,160613} & A2V & 0.086 & 95  & 389   & 98   & 61   & 323  &  &  &  &  & C & \\
	\object{HD\,161868} & A0V & 0.043 & 185 & 388.1 & 772  & 763  & 2860 &  &  &  &  & C & \\
	\object{HD\,164259} & F3V & 0.39  & 80  & 833.9 & 66   & 18   & 190  & 198 & 113 & 651 & G & V & \\
	\object{HD\,167468} & A0V & 0.043 & 295 & 278.2 & 968  & 482  & 3371 &  &  &  &  & V & \\
	\object{HD\,171834} & F3V & 0.386 & 60  & 388.1 & 22   & 19   & 69   & 170 & 125 & 612 & G & C & \\
	\object{HD\,172555} & A7V & 0.199 & 175 & 841.8 & 256  & 62   & 1165 &  &  &  &  & V & \\
	\object{HD\,175638} & A5V & 0.161 & 145 & 389.1 & 191  & 77   & 942  & 180186 & 1019 & 1247939 & B & V & \\
	\object{HD\,175639} & A5V & 0.204 & 200 & 110.8 & 735  & 257  & 2033 &  &  &  &  & V & \\
	\object{HD\,176638} & A0V & -0.021&260.8& 278.3 & 1998 & 1287 & 5542 &  &  &  &  & C & \\
	\object{HD\,177178} & A4V & 0.182 & 155 & 280.3 & 546  & 187  & 1932 &  &  &  &  & V & \\
	\object{HD\,177724} & A0V & 0.014 & 295 & 833.9 & 2752 & 1399 & 9102 &  &  &  &  & C & \\
	\object{HD\,177756} & B9V & -0.096& 160 & 833.8 & 5101 & 555  & 17662&  &  &  &  & V & V\\
	\object{HD\,181296} & A0V & 0.02  & 420 &1283.8 & 1559 & 905  & 6283 &  &  &  &  & C & \\
	\object{HD\,184985} & F7V & 0.501 & 5   & 389.1 & 3    & 1    & 13   & 2 & 1 & 7 & G & V & \\
	\object{HD\,186543} & A9V & 0.196 &121.5& 837   & 150  & 53   & 614  & 2137 & 583 & 15115 & B & V & \\
	\object{HD\,187532} & F0V & 0.402 & 95  & 829.9 & 33   & 21   & 133  & 240 & 143 & 974 & G & C & \\
	\object{HD\,188228} & A0V & -0.032& 115 & 841.9 & 117  & 90   & 450  &  &  &  &  & C & \\
	\object{HD\,189245} & F7V & 0.498 & 100 & 259.1 & 86   & 17   & 267  & 137 & 109 & 503 & G & V & \\
	\object{HD\,191862} & F5V & 0.476 & 8   & 364.2 & 3    & 2    & 12   & 9 & 4 & 31 & G & C & \\
	\object{HD\,196385} & A9V & 0.328 & 13  & 568.7 & 12   & 5    & 44   & 19 & 12 & 84 & G & V & \\
	\object{HD\,197692} & F5V & 0.426 & 40  & 829.9 & 30   & 7    & 118  & 85 & 33 & 307 & G & V & \\
	\object{HD\,198390} & F5V & 0.42  & 6.5 & 389.1 & 3    & 2    & 9    & 6 & 3 & 18 & G & C & \\
	\object{HD\,199254} & A4V & 0.131 & 145 & 368   & 178  & 93   & 637  & 2678 & 1270 & 10547 & B & C & \\
	\object{HD\,199260} & F7V & 0.507 & 13  & 829.9 & 13   & 2    & 50   & 22 & 6 & 74 & G & V & \\
	\object{HD\,200761} & A1V & -0.01 & 80  & 841.8 & 2095 & 107  & 5619 &  &  &  &  & V & V\\
	\object{HD\,202730} & A5V & 0.191 & 210 & 366.9 & 107  & 86   & 387  &  &  &  &  & C & \\
	\object{HD\,203608} & F6V & 0.494 & 8   & 909.8 & 2    & 1    & 6    & 2 & 1 & 9 & G & V & \\
	\object{HD\,205289} & F5V & 0.423 & 45  & 829.9 & 29   & 14   & 94   & 62 & 81 & 208 & G & C & \\
	\object{HD\,209819} & B8V & -0.075& 135 & 387.9 & 8182 & 220  & 21581&  &  &  &  & V & X\\
	\object{HD\,210302} & F6V & 0.489 & 12  &1287.7 & 9    & 2    & 38   & 10 & 4 & 40 & G & V & \\
	\object{HD\,210418} & A2V & 0.086 & 130 & 836.8 & 298  & 144  & 971  &  &  &  &  & V & \\
	\object{HD\,210739} & A3V & 0.169 & 160 & 837.9 & 375  & 307  & 1365 &  &  &  &  & C & \\
	\object{HD\,211976} & F6V & 0.451 & 5   & 368   & 4    & 2    & 11   & 4 & 3 & 16 & G & V & \\
	\object{HD\,212728} & A3V & 0.208 &254.9&1165.9 & 749  & 501  & 2351 &  &  &  &  & C & \\
	\object{HD\,213398} & A1V & 0.011 & 45  &1161.9 & 38   & 24   & 124  & 60 & 50 & 208 & G & C & \\
	\object{HD\,213845} & F7V & 0.446 & 25  & 841.8 & 26   & 6    & 99   & 40 & 25 & 156 & G & V & \\
	\object{HD\,215789} & A3V & 0.083 & 270 & 719   & 241  & 306  & 807  &  &  &  &  & C & \\
	\object{HD\,216627} & A3V & 0.066 & 70  & 389.2 & 4058 & 35   & 9514 & 154 & 274 & 519 & G & V & B\\
	\object{HD\,216956} & A3V & 0.145 & 85  & 325.2 & 52   & 17   & 282  & 277 & 150 & 1256 & G & V & \\
	\object{HD\,219482} & F7V & 0.521 & 7   & 837.9 & 14   & 2    & 38   & 10 & 3 & 36 & G & V & \\
	\object{HD\,220729} & F4V & 0.409 & 20  & 386   & 8018 & 4    & 19021& 6428 & 10 & 14987 & G & V & B\\
	\object{HD\,222095} & A2V & 0.082 & 165 & 387   & 200  & 109  & 741  &  &  &  &  & C & \\
	\object{HD\,222368} & F7V & 0.507 & 7   & 837.8 & 3    & 1    & 12   & 2 & 2 & 12 & G & V & \\
	\object{HD\,222603} & A7V & 0.2   & 60  & 385.1 & 296  & 20   & 765  & 573 & 98 & 1442 & G & V & \\
	\object{HD\,222661} & B9V & -0.032& 135 & 388.1 & 348  & 239  & 1031 &  &  &  &  & C & \\
	\object{HD\,223011} & A7V & 0.21  & 35.3& 840.9 & 364  & 13   & 1177 & 282 & 63 & 1056 & G & V & \\
	\object{HD\,223352} & A0V & 0.001 & 280 & 708.1 & 1643 & 1453 & 4917 &  &  &  &  & C & \\
	\object{HD\,223781} & A4V & 0.186 & 165 & 368   & 365  & 335  & 1148 &  &  &  &  & C & \\
	\object{HD\,224392} & A1V & 0.06  & 250 & 385.9 & 325  & 383  & 944  &  &  &  &  & C & \\
	\hline
      \end{longtable}
  \end{center}
  }
}

\longtab{3}{
\small{
\begin{center}
\begin{longtable}{l l l l l l l l l l}
\caption{Detection limits, either achievable or achieved with a 68.2$\%$ or a 99.7$\%$ probability in the present survey (see text). Note that only those detection limits less than 0.05 Jupiter mass are given with 2 digits.}
 \label{limdetsample}
 \\
\hline
HD & Achievable  & Achieved & Achieved  & Achievable & Achieved  & Achieved  & Achievable  & Achieved  & Achieved \\
   & P=3days     & P=3days  & P=3days   & P=10days   & P=10days  & P=10days  & P=100days   & P=100days & P=100days \\
   & (\Mjup)      & (\Mjup)   & (\Mjup)    & (\Mjup)     & (\Mjup)    & (\Mjup)    & (\Mjup)      & (\Mjup)    & (\Mjup)\\
   &             & P=68.2$\%$ & P=99.7 $\%$             &                     & P=68.2  $\%$    & P=99.7 $\%$          &       & P=68.2$\%$         & P=99.7$\%$ \\
\hline
693 & 0.02 &  &  & 0.04 &  &  & 0.1 &  & \\
2696 & 4.7 &  &  & 7.0 &  &  & 15.0 &  & \\
2834 &  &  &  &  &  &  &  &  & \\
2884 & 13.8 & 16.8 & 36.0 & 20.4 & 22.4 & 27.1 & 44.4 & 100.0 & 100.0\\
2885 &  &  &  &  &  &  &  &  & \\
3003 & 2.7 & 3.2 & 6.3 & 4.1 & 5.8 & 12.5 & 8.8 & 25.6 & 100.0\\
4247 & 0.4 & 0.5 & 1.0 & 0.6 & 0.9 & 1.8 & 1.2 & 2.0 & 6.3\\
4293 & 1.4 & 1.5 & 1.7 & 2.0 & 3.9 & 49.3 & 4.4 & 6.5 & 33.8\\
7439 & 0.1 &  &  & 0.2 &  &  & 0.4 &  & \\
9672 & 3.1 &  &  & 4.6 &  &  & 9.9 &  & \\
11262 &  &  &  &  &  &  &  &  & \\
12311 &  &  &  &  &  &  &  &  & \\
13555 & 0.2 & 0.2 & 0.4 & 0.3 & 0.5 & 8.2 & 0.6 & 3.4 & 100.0\\
14943 & 2.3 & 2.2 & 2.4 & 3.4 & 3.5 & 4.7 & 7.3 & 8.4 & 10.8\\
15008 & 7.5 & 9.2 & 15.2 & 11.2 & 14.8 & 36.2 & 24.3 & 32.2 & 91.4\\
17848 & 5.5 & 6.2 & 7.3 & 8.1 & 10.5 & 21.8 & 17.5 & 25.7 & 100.0\\
18978 & 3.1 & 6.4 & 10.6 & 4.7 & 9.1 & 62.9 & 10.1 & 21.3 & 44.4\\
19107 & 2.4 &  &  & 3.5 &  &  & 7.6 &  & \\
19545 &  &  &  &  &  &  &  &  & \\
21882 & 6.1 &  &  & 9.1 &  &  & 19.5 &  & \\
25457 & 0.4 & 0.4 & 0.4 & 0.5 & 0.6 & 0.6 & 1.2 & 2.1 & 31.8\\
25490 & 1.4 &  &  & 2.1 &  &  & 4.6 &  & \\
29488 & 3.2 & 3.1 & 3.2 & 4.7 & 4.9 & 5.1 & 10.3 & 10.3 & 11.6\\
29875 & 5.8 &  &  & 8.6 &  &  & 18.6 &  & \\
29992 & 4.7 & 9.2 & 100.0 & 7.0 & 12.5 & 100.0 & 15.2 & 21.8 & 48.4\\
30652 & 0.2 & 0.2 & 0.2 & 0.3 & 0.3 & 0.4 & 0.6 & 0.9 & 2.1\\
30739 & 20.0 & 23.2 & 44.9 & 29.6 & 44.9 & 94.2 & 64.4 & 81.9 & 100.0\\
31746 & 0.2 & 0.3 & 0.3 & 0.4 & 0.4 & 0.5 & 0.8 & 1.1 & 2.4\\
32743 & 0.2 & 0.2 & 0.2 & 0.3 & 0.4 & 1.2 & 0.6 & 1.1 & 100.0\\
32977 & 1.3 & 1.3 & 2.1 & 1.9 & 2.0 & 2.8 & 4.1 & 5.4 & 14.1\\
33256 & 0.05 & 0.1 & 0.1 & 0.1 & 0.1 & 0.2 & 0.2 & 0.4 & 100.0\\
33262 & 0.2 & 0.2 & 0.3 & 0.3 & 0.5 & 1.2 & 0.7 & 1.0 & 1.8\\
37306 & 5.5 & 5.3 & 6.4 & 8.1 & 8.4 & 9.0 & 17.7 & 21.9 & 46.8\\
38393 & 0.1 & 0.1 & 0.1 & 0.1 & 0.1 & 0.1 & 0.2 & 0.2 & 0.2\\
38678 & 13.4 & 13.2 & 16.8 & 19.8 & 21.9 & 29.3 & 43.0 & 57.5 & 97.6\\
39014 & 8.2 & 8.3 & 10.1 & 12.1 & 14.7 & 22.1 & 26.4 & 27.1 & 27.9\\
39060 & 4.5 & 4.4 & 4.7 & 6.6 & 7.7 & 9.2 & 14.4 & 17.3 & 29.3\\
40136 & 0.1 & 0.2 & 0.2 & 0.2 & 0.3 & 0.6 & 0.5 & 0.5 & 0.6\\
41695 & 11.2 &  &  & 16.7 &  &  & 35.9 &  & \\
41742 &  &  &  &  &  &  &  &  & \\
43940 & 10.5 & 16.9 & 100.0 & 15.6 & 16.5 & 26.6 & 33.8 & 67.2 & 100.0\\
46089 & 16.5 &  &  & 24.6 &  &  & 53.0 &  & \\
49095 & 0.04 & 0.04 & 0.04 & 0.1 & 0.1 & 0.1 & 0.1 & 0.2 & 0.3\\
49933 & 0.4 &  &  & 0.6 &  &  & 1.2 &  & \\
50445 & 1.1 & 1.1 & 1.5 & 1.6 & 1.8 & 2.1 & 3.5 & 3.7 & 4.1\\
54834 &  &  &  &  &  &  &  &  & \\
56537 & 3.3 & 3.9 & 5.3 & 4.9 & 5.9 & 7.8 & 10.6 & 14.8 & 29.5\\
59984 & 0.05 & 0.1 & 0.3 & 0.1 & 0.1 & 3.5 & 0.1 & 0.3 & 100.0\\
60532 & 0.4 & 0.4 & 0.4 & 0.6 & 0.6 & 0.7 & 1.3 & 1.4 & 1.9\\
60584 & 0.3 & 0.3 & 0.4 & 0.5 & 0.6 & 0.8 & 1.0 & 1.9 & 4.0\\
63847 & 11.8 & 11.9 & 12.8 & 17.4 & 23.0 & 51.7 & 37.8 & 49.2 & 88.1\\
66664 & 11.4 & 11.6 & 16.1 & 16.9 & 19.8 & 25.6 & 36.6 & 75.4 & 100.0\\
68146 & 0.1 & 0.1 & 0.1 & 0.1 & 0.1 & 0.1 & 0.2 & 0.2 & 0.2\\
68456 &  &  &  &  &  &  &  &  & \\
71155 & 10.5 & 10.5 & 11.4 & 15.6 & 17.4 & 21.0 & 33.9 & 53.0 & 100.0\\
73262 & 22.0 &  &  & 32.8 &  &  & 70.7 &  & \\
74591 & 2.8 & 2.9 & 3.1 & 4.1 & 4.4 & 4.8 & 8.9 & 14.1 & 34.0\\
74873 & 26.0 &  &  & 38.8 &  &  & 83.5 &  & \\
75171 & 4.4 &  &  & 6.6 &  &  & 14.1 &  & \\
76653 & 0.2 & 0.2 & 0.3 & 0.3 & 0.3 & 0.8 & 0.6 & 1.0 & 1.6\\
77370 & 0.5 & 0.6 & 0.7 & 0.8 & 1.1 & 2.8 & 1.7 & 2.2 & 5.2\\
82165 & 8.7 & 18.7 & 100.0 & 12.9 & 14.5 & 17.6 & 27.9 & 71.1 & 100.0\\
83446 & 4.6 & 6.0 & 16.3 & 6.8 & 6.8 & 7.6 & 14.8 & 20.1 & 48.1\\
88955 & 2.3 & 2.3 & 2.3 & 3.4 & 4.9 & 10.0 & 7.4 & 18.8 & 100.0\\
89328 & 5.0 & 5.2 & 6.3 & 7.4 & 10.2 & 39.9 & 16.0 & 18.8 & 32.5\\
90132 & 6.7 & 11.4 & 100.0 & 10.0 & 18.7 & 100.0 & 21.6 & 71.1 & 100.0\\
91324 & 0.04 & 0.05 & 0.1 & 0.1 & 0.1 & 0.1 & 0.1 & 0.2 & 0.5\\
91889 & 0.1 & 0.1 & 0.1 & 0.1 & 0.1 & 0.3 & 0.2 & 0.2 & 0.4\\
93372 & 0.1 & 0.1 & 0.2 & 0.1 & 0.1 & 0.8 & 0.2 & 0.2 & 0.2\\
94388 & 0.2 & 0.2 & 0.2 & 0.3 & 0.6 & 1.6 & 0.7 & 1.7 & 100.0\\
96819 & 8.4 &  &  & 12.5 &  &  & 27.0 &  & \\
97244 & 1.4 & 1.6 & 2.2 & 2.0 & 2.1 & 2.7 & 4.4 & 12.0 & 100.0\\
97603 & 5.6 & 8.0 & 32.8 & 8.3 & 11.1 & 20.0 & 18.1 & 41.2 & 100.0\\
99211 & 2.2 & 3.5 & 15.5 & 3.2 & 6.6 & 100.0 & 7.0 & 19.9 & 100.0\\
99453 &  &  &  &  &  &  &  &  & \\
100563 & 0.04 &  &  & 0.1 &  &  & 0.1 &  & \\
101198 & 0.5 & 0.5 & 0.6 & 0.8 & 0.9 & 1.7 & 1.7 & 2.6 & 20.3\\
102124 & 5.0 & 10.2 & 100.0 & 7.4 & 18.3 & 100.0 & 16.1 & 52.4 & 100.0\\
102647 & 2.1 & 2.1 & 2.6 & 3.1 & 3.3 & 4.2 & 6.7 & 18.1 & 100.0\\
104731 & 0.3 & 0.7 & 9.5 & 0.5 & 1.2 & 100.0 & 1.0 & 3.5 & 100.0\\
105850 & 3.6 & 4.0 & 6.3 & 5.3 & 8.5 & 100.0 & 11.6 & 29.2 & 100.0\\
106661 & 7.0 & 7.4 & 19.7 & 10.4 & 26.0 & 100.0 & 22.5 & 39.9 & 100.0\\
109085 & 0.3 & 0.3 & 0.4 & 0.4 & 0.6 & 6.1 & 1.0 & 1.1 & 2.1\\
109787 & 16.6 & 26.9 & 100.0 & 24.7 & 51.9 & 100.0 & 53.5 & 100.0 & 100.0\\
110411 & 16.4 & 18.4 & 21.4 & 24.3 & 34.0 & 54.0 & 52.8 & 100.0 & 100.0\\
111998 & 0.5 & 0.6 & 0.9 & 0.8 & 1.3 & 8.7 & 1.7 & 2.0 & 3.1\\
112934 &  &  &  &  &  &  &  &  & \\
114642 & 0.7 & 0.8 & 2.0 & 1.0 & 1.3 & 1.7 & 2.1 & 35.5 & 100.0\\
115892 & 1.1 & 1.2 & 1.8 & 1.7 & 2.3 & 3.8 & 3.7 & 67.6 & 100.0\\
116160 &  &  &  &  &  &  &  &  & \\
116568 &  &  &  &  &  &  &  &  & \\
118098 & 7.2 & 6.7 & 7.8 & 10.7 & 11.6 & 18.3 & 23.1 & 65.4 & 100.0\\
124850 & 0.5 & 0.6 & 0.9 & 0.8 & 1.0 & 1.8 & 1.6 & 30.2 & 100.0\\
125276 & 0.02 & 0.03 & 0.7 & 0.02 & 0.03 & 0.1 & 0.1 & 0.1 & 100.0\\
126248 & 10.8 & 12.0 & 14.4 & 16.0 & 17.3 & 22.4 & 34.7 & 87.7 & 100.0\\
128020 & 0.02 & 0.03 & 0.03 & 0.05 & 0.1 & 0.5 & 0.1 & 0.1 & 100.0\\
128167 & 0.3 & 0.3 & 0.5 & 0.4 & 0.5 & 0.9 & 1.0 & 14.4 & 100.0\\
128898 & 0.9 &  &  & 1.4 &  &  & 3.0 &  & \\
129422 & 7.7 &  &  & 11.5 &  &  & 24.7 &  & \\
129926 & 5.1 & 6.3 & 10.1 & 7.6 & 10.8 & 40.2 & 16.5 & 38.9 & 100.0\\
130109 & 26.6 &  &  & 39.8 &  &  & 85.6 &  & \\
132052 & 1.9 &  &  & 2.8 &  &  & 6.0 &  & \\
133469 & 0.2 &  &  & 0.3 &  &  & 0.7 &  & \\
135379 & 0.6 & 1.1 & 12.1 & 0.9 & 1.5 & 8.2 & 1.9 & 2.5 & 6.9\\
135559 & 8.4 & 8.4 & 9.2 & 12.5 & 13.0 & 15.9 & 27.0 & 70.8 & 100.0\\
138763 & 0.7 & 0.7 & 0.8 & 1.0 & 1.3 & 2.0 & 2.2 & 3.9 & 19.8\\
139211 & 0.04 & 0.04 & 0.05 & 0.1 & 0.1 & 1.7 & 0.1 & 0.2 & 100.0\\
141513 & 3.3 & 3.4 & 4.1 & 4.9 & 6.0 & 7.8 & 10.7 & 25.4 & 100.0\\
141851 & 10.0 &  &  & 15.0 &  &  & 32.3 &  & \\
142139 & 0.8 &  &  & 1.2 &  &  & 2.5 &  & \\
142629 &  &  &  &  &  &  &  &  & \\
145689 & 3.2 &  &  & 4.8 &  &  & 10.2 &  & \\
146514 & 11.0 &  &  & 16.5 &  &  & 35.4 &  & \\
146624 & 0.2 &  &  & 0.3 &  &  & 0.7 &  & \\
147449 & 1.0 & 1.0 & 1.1 & 1.4 & 1.5 & 2.0 & 3.1 & 6.1 & 57.2\\
153363 & 2.8 & 3.7 & 8.9 & 4.2 & 16.1 & 100.0 & 9.1 & 13.2 & 24.9\\
156751 & 6.8 &  &  & 10.2 &  &  & 22.0 &  & \\
158094 &  &  &  &  &  &  &  &  & \\
158352 &  &  &  &  &  &  &  &  & \\
159170 & 9.9 &  &  & 14.7 &  &  & 31.7 &  & \\
159492 & 1.7 & 3.4 & 100.0 & 2.6 & 3.8 & 18.3 & 5.6 & 10.7 & 100.0\\
160613 & 1.7 &  &  & 2.6 &  &  & 5.6 &  & \\
161868 & 14.5 &  &  & 21.7 &  &  & 46.8 &  & \\
164259 & 0.9 & 1.2 & 2.4 & 1.4 & 2.0 & 3.6 & 3.0 & 4.2 & 8.2\\
167468 & 19.6 & 23.4 & 31.6 & 29.1 & 46.4 & 100.0 & 63.1 & 83.4 & 100.0\\
171834 & 0.3 & 0.4 & 0.5 & 0.5 & 0.5 & 0.7 & 1.0 & 2.2 & 100.0\\
172555 & 4.2 & 5.1 & 11.3 & 6.2 & 7.6 & 11.3 & 13.5 & 20.7 & 33.0\\
175638 & 3.3 & 3.8 & 4.7 & 4.8 & 5.9 & 8.0 & 10.5 & 17.8 & 38.3\\
175639 & 11.2 &  &  & 16.7 &  &  & 35.9 &  & \\
176638 & 41.2 &  &  & 61.6 &  &  & 132.7 &  & \\
177178 & 8.5 &  &  & 12.7 &  &  & 27.4 &  & \\
177724 & 54.0 &  &  & 80.7 &  &  & 173.9 &  & \\
177756 &  &  &  &  &  &  &  &  & \\
181296 & 32.7 & 40.0 & 76.3 & 48.4 & 49.3 & 68.7 & 105.2 & 100.0 & 100.0\\
184985 & 0.04 & 0.1 & 0.1 & 0.1 & 0.1 & 0.1 & 0.1 & 0.2 & 0.6\\
186543 & 2.5 & 2.6 & 3.3 & 3.6 & 4.2 & 6.9 & 7.9 & 12.3 & 43.9\\
187532 & 0.5 & 0.6 & 1.1 & 0.7 & 0.7 & 0.8 & 1.5 & 3.9 & 10.4\\
188228 & 2.6 & 3.2 & 6.7 & 3.9 & 5.7 & 72.4 & 8.5 & 10.0 & 17.6\\
189245 & 1.1 & 2.7 & 100.0 & 1.7 & 2.4 & 10.0 & 3.6 & 8.2 & 100.0\\
191862 & 0.04 &  &  & 0.1 &  &  & 0.1 &  & \\
196385 & 0.2 & 0.2 & 0.4 & 0.3 & 0.3 & 1.1 & 0.6 & 1.0 & 100.0\\
197692 & 0.4 & 0.4 & 0.5 & 0.6 & 0.7 & 0.7 & 1.3 & 9.4 & 100.0\\
198390 & 0.04 &  &  & 0.1 &  &  & 0.1 &  & \\
199254 & 3.2 & 4.7 & 7.0 & 4.7 & 4.9 & 4.9 & 10.2 & 16.9 & 100.0\\
199260 & 0.2 & 0.2 & 0.3 & 0.2 & 0.3 & 0.4 & 0.5 & 1.2 & 13.8\\
200761 &  &  &  &  &  &  &  &  & \\
202730 & 1.7 &  &  & 2.5 &  &  & 5.3 &  & \\
203608 & 0.02 & 0.02 & 0.03 & 0.03 & 0.04 & 0.1 & 0.1 & 0.2 & 100.0\\
205289 & 0.4 & 0.5 & 0.8 & 0.6 & 0.6 & 0.8 & 1.3 & 6.1 & 100.0\\
209819 &  &  &  &  &  &  &  &  & \\
210302 & 0.1 & 0.1 & 0.1 & 0.2 & 0.2 & 0.2 & 0.4 & 0.6 & 1.0\\
210418 & 5.7 & 7.0 & 13.6 & 8.4 & 11.9 & 38.7 & 18.2 & 21.4 & 28.8\\
210739 & 5.9 &  &  & 8.9 &  &  & 19.1 &  & \\
211976 & 0.1 & 0.1 & 100.0 & 0.1 & 0.1 & 0.2 & 0.2 & 0.3 & 100.0\\
212728 & 12.4 & 13.0 & 15.5 & 18.3 & 28.2 & 60.8 & 39.8 & 63.9 & 100.0\\
213398 & 0.8 & 1.4 & 8.5 & 1.2 & 1.7 & 5.9 & 2.6 & 3.9 & 9.4\\
213845 & 0.3 & 0.3 & 0.4 & 0.5 & 0.6 & 0.8 & 1.1 & 1.6 & 3.5\\
215789 & 4.6 & 3.5 & 100.0 & 6.8 & 4.7 & 7.1 & 14.8 & 3.4 & 100.0\\
216627 &  &  &  &  &  &  &  &  & \\
216956 & 0.9 & 3.6 & 100.0 & 1.3 & 3.2 & 8.1 & 2.9 & 10.0 & 100.0\\
219482 & 0.2 & 0.2 & 0.4 & 0.3 & 0.4 & 2.6 & 0.6 & 0.9 & 2.4\\
220729 &  &  &  &  &  &  &  &  & \\
222095 & 3.8 & 4.1 & 5.5 & 5.7 & 7.6 & 11.0 & 12.3 & 15.9 & 39.9\\
222368 & 0.04 & 0.05 & 0.1 & 0.1 & 0.1 & 0.1 & 0.1 & 0.1 & 0.2\\
222603 & 4.5 &  &  & 6.8 &  &  & 14.6 &  & \\
222661 & 7.3 &  &  & 10.9 &  &  & 23.5 &  & \\
223011 & 6.0 & 6.7 & 9.6 & 8.9 & 13.2 & 96.8 & 19.3 & 30.4 & 79.8\\
223352 & 32.9 &  &  & 49.1 &  &  & 105.8 &  & \\
223781 & 5.7 &  &  & 8.5 &  &  & 18.2 &  & \\
224392 & 6.0 &  &  & 8.9 &  &  & 19.2 &  & \\
\end{longtable}
  \end{center}
}
}

\end{document}